\documentclass[journal]{IEEEtran}
\usepackage{amsmath}
\usepackage{amssymb}
\usepackage{amsthm}
\usepackage{cite}
\usepackage{color}
\usepackage{xcolor}
\usepackage{epsfig,latexsym}
\usepackage{float}
\usepackage{fancyhdr}
\usepackage{graphicx}
\usepackage{indentfirst}
\usepackage{mdwmath}
\usepackage{mdwtab}
\usepackage{setspace}
\usepackage{times}
\usepackage{url}
\usepackage{algorithm,algorithmic}

\newtheorem{remark}{Remark}
\newtheorem{theorem}{Theorem}

\newtheorem{lemma}{Lemma}

\newtheorem{corollary}{Corollary}

\newtheorem{proposition}{Proposition}

\begin{document}

\title{On the Blockage Effect \\ in Pinching-Antenna Systems (PASS)}
\author{Jinhua Wang, Jun Wang, Tianwei Hou,~\IEEEmembership{Member,~IEEE,} Xin Sun, \\ and Arumugam Nallanathan,~\IEEEmembership{Fellow,~IEEE}

\thanks{This work is supported in part by the Beijing Natural Science Foundation L232041, and in part by the EPSRC grant numbers to acknowledge are EP/W004100/1, EP/W034786/1, EP/Y037243/1 and EP/X04047X/2. (Corresponding author: Tianwei Hou.)}
\thanks{Jinhua Wang, Jun Wang, and Xin Sun are with the School of Electronic and Information Engineering, Beijing Jiaotong University, Beijing 100044, China (e-mail: 25110041@bjtu.edu.cn, wangjun1@bjtu.edu.cn, and xsun@bjtu.edu.cn).}
\thanks{Tianwei Hou is with the Beijing Key Laboratory of Convergent Communications and Networking Technologies in LEO Satellite Systems, and also with the State Key Laboratory of Networking and Switching Technology, Beijing University of Posts and Telecommunications, Beijing 100876, China (email: htw@bupt.edu.cn).}
\thanks{Arumugam Nallanathan is with the School of Electronic Engineering and Computer Science, Queen Mary University of London, London E1 4NS, U.K., and also with the Department of Electronic Engineering, Kyung Hee University, Yongin-si, Gyeonggi-do 17104, Korea. (e-mail: a.nallanathan@qmul.ac.uk).}
}

\maketitle

\begin{abstract}
Pinching-antenna systems (PASS) offer considerable potential for wireless communications due to their unique ability to dynamically reconfigure radiation positions along a waveguide. However, the performance of PASS remains a critical challenge in the presence of random line-of-sight (LoS) blockage, leading to significant attenuation and even communication outages. In this paper, the performance of PASS in the presence of LoS blockage is investigated from the perspective of stochastic geometry. Obstacles are modeled through a homogeneous Poisson point process (PPP), where the geometric dimensions, numbers, and positions are treated as random variables. To conduct a concrete characterization of LoS blockage, the random-height-and-random-radius (RHRR) blockage model and the deterministic-height-and-deterministic-radius (DHDR) blockage model are proposed. In particular, closed-form analytical and asymptotic expressions for the outage probability are obtained, along with analytical and approximate expressions for the ergodic rate. Our simulation results reveal that denser obstacle environments or statistically larger obstacles substantially increase the probability of LoS blockage and degrade the system performance. Moreover, owing to its ability to dynamically reposition PAs, PASS can consistently outperform conventional antenna systems in the presence of LoS blockage.
\end{abstract}


\section{Introduction}\label{sec:Intro}

Flexible antenna technologies have attracted significant attention as a promising approach to enhancing the reliability and adaptability of the sixth generation (6G) wireless communication systems~\cite{10858129}. Unlike conventional fixed-position antennas, which are vulnerable to performance degradation under time-varying propagation conditions, flexible antenna architectures offer additional diversity and multiplexing gains. Reconfigurable intelligent surfaces (RIS) provide a means to create equivalent line-of-sight (LoS) links by dynamically reflecting or transmitting signals to bypass obstacles~\cite{9140329, 9424177}. Nevertheless, the effectiveness of RIS is limited by multiplicative fading across the cascaded channels and its dependence on accurate phase alignment. Fluid antennas and movable antennas enhance spatial diversity by physically or electronically shifting antenna positions within a small region, thereby partially mitigating random fading and transient blockages~\cite{10286328, 9264694}. However, since their repositioning ranges are typically limited to just a few wavelengths, they remain insufficient to against large-scale or persistent blockages.

Pinching-antenna systems (PASS) have recently been proposed as an innovative antenna architecture, which extends the concept of flexible antennas by enabling the large-scale reconfiguration of radiation points along the length of a dielectric waveguide~\cite{suzuki2022pinching, example2025}. PASS integrates multiple pinching-antennas (PAs) onto a waveguide, and its key advantage lies in the capability to dynamically reposition the effective PAs across a wide spatial range, thereby enabling adaptive beamforming and spatial diversity gains for user equipment (UE) located in complex environments~\cite{11169486, yang2025pinchingantennasprinciplesapplications}. By flexibly shifting the radiation points, PASS can establish favorable LoS links without relying on external reflectors or purely passive structures. Therefore, compared with other flexible antenna technologies, PASS offers additional performance gains, representing a promising pathway for achieving reliable high-frequency communications~\cite{ding2025flexible}.

Recent contributions on PASS have expanded across multiple domains. The resource allocation scheme, involving the optimal positions and number of activated PAs, is fundamental for maximizing system throughput~\cite{10909665, 10912473}. The beamforming and optimization principles of PAs are further investigated in~\cite{wang2025modelingbeamformingoptimizationpinchingantenna}, where a physics-based coupler model for PASS is proposed. The downlink rate maximization is explored in~\cite{10896748}, which solves the joint problem of minimizing large-scale path loss and managing dual phase shifts from both waveguide and free-space propagation. The performance of the uplink PASS is rigorously analyzed in~\cite{11195810}, which demonstrates that an asymmetric non-uniform distribution of PA positions can significantly enhance the achievable sum rate in a multiple-PA scenario. In addition, the application space of PASS is broadened to include integrated sensing and communication~\cite{11197530, li2025pinchingantennaintegratedsensing}, indoor positioning~\cite{zhang2025pinchingantennasystemspassbasedindoor}, simultaneous wireless information and power transfer~\cite{11106459}, among others. However, due to its waveguide-based architecture and deployment patterns, PASS is inherently susceptible to LoS blockage caused by surrounding obstacles whether in outdoor or indoor scenarios.

LoS blockage is widely recognized as a critical factor that severely degrades wireless system performance and link reliability, which originates from objects such as buildings, vehicles, and foliage in outdoor environments, as well as human bodies, walls, and furniture indoors. The impact of LoS blockage becomes increasingly significant at higher operating frequencies, owing to the fundamental propagation differences across frequency bands. In wireless fidelity and sub-6 GHz cellular systems, lower-frequency signals have stronger diffraction and penetration capabilities, which allows them to bend around or pass through moderate obstacles, typically incurring less than 10 dB of additional attenuation~\cite{10121509}. In contrast, the communication paradigms envisioned for 6G networks, which utilize millimeter-wave (mmWave) bands and even higher terahertz (THz) frequencies, are extremely susceptible to LoS blockage~\cite{8869705, 9512383}. While mmWave and higher bands offer vast bandwidths and extremely high data rates, their short wavelengths result in weak diffraction and strong absorption. Consequently, even small objects, such as human bodies or foliage, can cause severe power strength loss of 20–30 dB or more, leading to sudden and complete link outages~\cite{7593259}. Given that PASS operates primarily in such high-frequency regimes, its performance is particularly sensitive to LoS blockage, which poses a major challenge to achieving the stringent reliability and ultra-low latency requirements of next-generation networks. Consequently, a systematic analysis of the impact of LoS blockage on PASS is imperative, as it directly affects the design, optimization, and practical deployment of PASS across emerging application domains.

Several studies examine the impact of LoS blockage on PASS and provide unique insights~\cite{11036558, 11178241, xu2025pinchingantennadesignlosblockage}. While LoS blockage degrades performance in single-user scenarios, it may actually benefit multi-user systems by suppressing co-channel interference~\cite{11036558}. To exploit LoS blockage for interference suppression, the problem of joint waveguide and PA association is studied in~\cite{11178241}, where the size and position of obstacles are perfectly known. In~\cite{xu2025pinchingantennadesignlosblockage}, the in-waveguide attenuation is demonstrated to have negligible impact compared to blockage under meter-level propagation, particularly in densely obstructed environments. It is worth noting that the above mentioned contributions adopt simplified blockage models for analytical tractability, which mathematically quantify the severity of LoS blockage. For instance, the LoS probability is typically expressed as a simple exponential function parameterized by a LoS blockage exponent~\cite{3gpp.38.901}. Although the blockage models built by mathematical fitting are convenient for revealing general trends in system-level analysis, they weaken the randomness of blockage effects, which limits the physical interpretability on blockage caused by real-world obstacles and restricts the applicability in more complex or dynamic scenarios.

Stochastic geometry is a powerful tool for modelling the spatial configuration of wireless networks, which represents the network entities such as base stations (BSs), UEs, and obstacles as realizations of random point processes. By providing a probabilistic representation, stochastic geometry bridges the gap between physical modelling and analytical tractability, revealing the influence of fundamental spatial parameters, such as the spatial density and the path-loss exponent, on the system performance~\cite{9516701}.
The application of stochastic geometry to characterize blockage effects is well-established in mmWave and THz systems, including urban building blockage~\cite{6840343, 9253591}, human-body blockage~\cite{human_blockage1}, dynamic blockage~\cite{8643739}, indoor blockage~\cite{9247469}, and so on. Collectively, these studies reveal the complexities of wireless environments involving dynamic UEs, irregularly shaped obstacles, and randomly distributed obstacles, demonstrating that stochastic geometry is a well-suited approach for modeling and analyzing LoS blockage in PASS.

\subsection{Motivation and Contribution}\label{subsec:MaC}

Although the concept of PASS has recently attracted considerable attention, research on PASS under LoS blockage remains rather limited. The few existing studies that address LoS blockage effects mainly rely on simplified or empirical assumptions, where the intrinsic randomness of practical obstacles is inevitably weakened~\cite{11036558, 11178241, xu2025pinchingantennadesignlosblockage}. Motivated by this gap, our work aims to investigate the performance of PASS in the presence of random LoS blockage, where obstacles are spatially and geometrically random. However, introducing randomness brings several fundamental challenges: i) A tractable obstacle modelling approach for PASS that balances physical interpretability and analytical conciseness has not been well established; ii) A systematic analytical treatment that incorporates random LoS blockage into the performance evaluation of PASS is currently absent; iii) The quantitative evolution of the performance gain achieved by PASS under random LoS blockage, particularly its dependence on obstacle intensity, geometry, and service area, remains largely unexplored.

To tackle the aforementioned challenges, a stochastic geometry approach is utilized to model random LoS blockage systematically, enabling the analytical evaluation on the performance of PASS in the obstructed environments. The key contributions of this work are summarized as follows:

\begin{itemize}
   \item To characterize the LoS blockage effect in PASS, obstacles are modeled by using a homogeneous Poisson point process (PPP), which captures the random number and spatial distribution of obstacles within the service area. Specifically, we propose the random-height-and-random-radius (RHRR) model with Gaussian-distributed heights and uniformly-distributed radii, alongside the deterministic-height-and-deterministic-radius (DHDR) model serving as a simplified counterpart by assuming fixed obstacle dimensions.
  \item Based on the proposed blockage models, closed-form expressions for the LoS probability are derived. Subsequently, closed-form analytical and asymptotic expressions for the outage probability are derived. Our analysis reveals that the outage probability is positively correlated not only with the spatial distribution and geometric dimensions of obstacles, but also with the area width.
  \item The analytical and approximate expressions for the ergodic rate are derived. Our analysis reveals that the ergodic rate is negatively correlated with both the spatial density of obstacles and the area width.
  \item Simulation results validate the theoretical derivations and yield several important insights: 1) Denser or statistically larger obstacles substantially increase outage probability and reduce ergodic rate; 2) In the high-SNR regime with LoS blockage, further increasing transmit power yields almost no improvement in outage probability for PASS, i.e., the diversity order remains zero, whereas the ergodic rate exhibits a linear growth; 3) Owing to its ability to dynamically reposition PAs, PASS can maintain higher performance over conventional antenna systems (CASS) in the presence of LoS blockage.
\end{itemize}

\subsection{Organization and Notations}\label{subsec:OaN}

The rest of the paper is organized as follows. The antenna model and the fundamental principles of the blockage model are introduced in Section~\ref{sec:SysModel}. The analytical results are presented in Section~\ref{sec:Analysis} to evaluate the performance of PASS with LoS blockage. The numerical results are demonstrated in Section~\ref{sec:Sim} to verify our analysis, which is concluded in Section~\ref{sec:Concl}. $\mathbb{E}\left(  \cdot  \right)$ denotes the mathematical expectation, and $\mathbb{P}\left(  \cdot  \right)$ denotes the probability. $\left|  \cdot  \right|$ denotes the modulus, and $\left\|  \cdot  \right\|$ denotes the Euclidean norm. $X \sim \mathcal{N}\left( \mu, \sigma^2 \right)$ denotes a Gaussian distribution with a mean of $\mu$ and a variance of $\sigma^2$. $ X \sim \mathcal{U} \left( a, b \right)$ denotes a continuous uniform distribution on the interval $\left[ {a,b} \right]$.

\begin{figure}[t!]
  \centering
  \includegraphics[width=3.3 in]{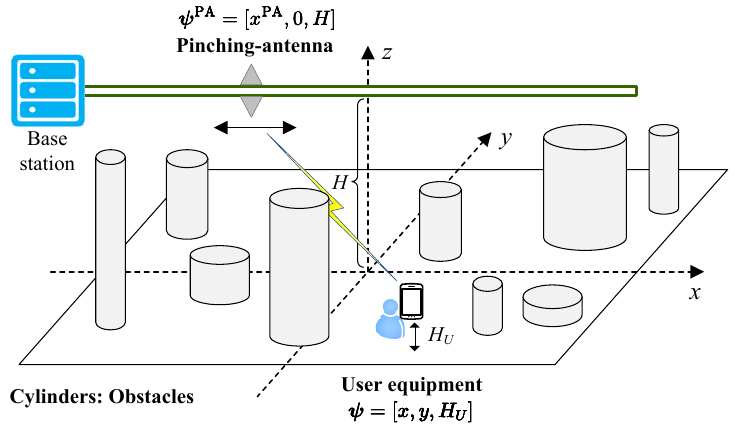}
  \caption{Illustration of PASS with obstacles.}\label{fig:LoS Blockage PASS model}
\end{figure}

\begin{figure}[t!]
    \centering
    \begin{minipage}{0.48\textwidth}
        \centering
        \includegraphics[width = 0.96\linewidth]{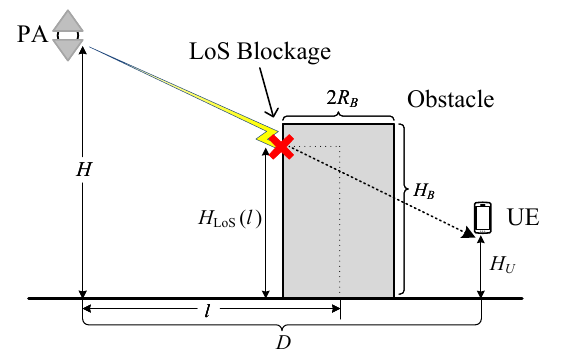}        
        \centerline{\footnotesize(a)}
    \end{minipage}
    \\[1em]
    \begin{minipage}{0.48\textwidth}
        \centering
        \includegraphics[width = 0.96\linewidth]{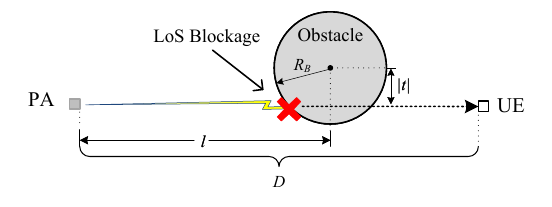}        
        \centerline{\footnotesize(b)}
    \end{minipage}
    \caption{Illustrations of LoS blockage on the LoS path between the PA and the UE: (a) Vertical view of the LoS path; (b) Top view of the LoS path.}
    \label{fig:LoS Path}
\end{figure}

\section{System Model}\label{sec:SysModel}
\subsection{Antenna Model}\label{subsec:AntModel}

As illustrated in Fig.~\ref{fig:LoS Blockage PASS model}, a rectangular service area with a length of $L$ and a width of $W$ is considered, which is denoted by $\mathcal{A}$. The center of the service area serves as the origin of the three-dimensional Cartesian coordinate system. A downlink PASS equipped with a single waveguide is considered. The waveguide is placed parallel to the $x$-axis, with its location defined as $[(-L/2,L/2),0,H]$, where $H$ denotes the height of the waveguide. The BS is positioned at $\boldsymbol{\psi} _0=[-L/2,0,H]$. To facilitate the performance analysis, the single-antenna UEs are randomly positioned in the service area, following a uniform spatial distribution. The PAs are activated on the waveguide at the location closest to the UEs in order to maximize the channel gain. The locations of the UE and the PA are denoted by $\boldsymbol{\psi} = [x,y,H_U]$ and $\boldsymbol{\psi} ^{\mathrm{PA}} = [x^{\mathrm{PA}},0,H]$, respectively, where $H_U$ denotes the height of the UE. The Euclidean distance between the PA and the UE is given by
\begin{equation}\label{distance_PA_to_UE}
\left\| {\boldsymbol{\psi} ^{\mathrm{PA}} - {\boldsymbol{\psi}}} \right\| = \sqrt {{\left( {x^{\mathrm{PA}} - x} \right)}^2 + y^2 + {\left( {H - {H_U}} \right)}^2}.
\end{equation}

From a PA to a UE, the large-scale fading governs the LoS channel gain. The small-scale fading is limited to the phase shift accumulated during propagation through the waveguide and free-space environment. For simplicity, the in-waveguide attenuation is omitted, since it has a negligible impact on the signal power during meter-level propagation~\cite{xu2025pinchingantennadesignlosblockage}. The LoS channel gain can be expressed as follows:
\begin{equation}\label{LoS_channel_gain}
h = \frac{c}{{4\pi {f_c}\left\| {{\boldsymbol{\psi} ^{\mathrm{PA}}} - \boldsymbol{\psi} } \right\|}}{e ^{ - j2\pi \left( {\frac{1}{\lambda_w }\left\| {\boldsymbol{\psi} ^{\mathrm{PA}} - \boldsymbol{\psi} } \right\| + \frac{1}{{\lambda _g}}\left\| {\boldsymbol{\psi} _0 - {\boldsymbol{\psi} ^{\mathrm{PA}}}} \right\|} \right)}},
\end{equation}
where $c$ denotes the speed of light, $f_c$ denotes the carrier frequency, $\lambda_w$ denotes the carrier wavelength, $\lambda _g = \lambda_w / n_{\mathrm{eff}}$ denotes waveguide wavelength with $n_{\mathrm{eff}}$ being the effective refractive index of a dielectric waveguide.

The binary discrete variable $\kappa$ is defined as the LoS indicator, representing whether the LoS path is blocked or not~\cite{11036558}. If there is LoS blockage between the PA and the UE, $\kappa =0$, otherwise, $\kappa =1$. The effective LoS channel gain takes the form as follows:
\begin{equation}\label{effective_LoS_channel}
\tilde h = \kappa h
\end{equation}

Accordingly, the signal received at the UE is formulated as follows:
\begin{equation}\label{UE_receive_signal}
r = \tilde h\sqrt P s + n,
\end{equation}
where $P$ represents the overall transmit power budget, $s$ represents the transmit signal of the UE with $\mathbb{E} (s)=0$, and $n \sim \mathcal{N}(0,\sigma^2)$ denotes the additive white Gaussian noise with zero mean and variance $\sigma^2$.

\subsection{Blockage Model}\label{subsec:BlkModel}

We aim to analyze the impact of LoS blockage on PASS caused by the spatial distribution and geometry characteristics of obstacles. To avoid analysis difficulties caused by differences in obstacle shapes, we model the obstacles as cylinders, with effective radius and height denoted by $R_B$ and $H_B$, respectively~\cite{human_blockage1, 9247469}. The projection of the LoS path in the $x$-$y$ plane, which is also the horizontal distance from the PA to the UE, is denoted by $D = \sqrt {{{\left( {{x^{{\mathrm{PA}}}} - x} \right)}^2} + {y^2}}$.

Our blockage model is introduced in terms of vertical and top views. In the vertical view, according to the geometric relationships as illustrated in Fig.~\ref{fig:LoS Path}(a), the vertical height of the LoS path can be regarded as a linear function of the horizontal propagation distance, which can be expressed as follows:
\begin{equation}\label{H_LoS_function}
H_{\mathrm{LoS}}\left( l \right) =  - \frac{H - H_U}{D}l + H,\quad 0 < l < D,
\end{equation}
where $l$ denotes the horizontal propagation distance of signal in the $x$-$y$ plane. In the top view, any obstacles blocking the LoS path must be close to it and sufficiently ``fat'', as illustrated in Fig.~\ref{fig:LoS Path}(b). Let $t$ denotes the signed horizontal distance from the obstacle center to the LoS path, where its sign indicates the side of the LoS path on which the obstacle is located. Hence, $\left| t \right|$ denotes the absolute horizontal distance.

For an effective LoS blockage event to occur, both the height and location conditions must be met simultaneously, which means that (i) the obstacle's height is greater than or equal to the LoS path's height, i.e., $H_B \geqslant H_{\mathrm{LoS}}\left( l \right)$, as shown in Fig.~\ref{fig:LoS Path}(a), and (ii) the distance from the center of the obstacle to the LoS path is less than or equal to the radius, i.e., $\left| t \right| < R_B$, as shown in Fig.~\ref{fig:LoS Path}(b).
In addition, these two conditions are independent of each other.

We model the random distribution of obstacles in the service area as a two-dimensional (2-D) homogeneous PPP with the intensity of $\lambda$. The number of obstacles, $N_B$, is a random variable that obeys a Poisson distribution with the intensity of $\lambda LW$. Since the blockage area is a subregion of the service area, the LoS blockage process is the independent thinning of the original 2-D homogeneous PPP of obstacle positions, which is also a PPP~\cite{chiu2013stochastic}. The intensity of the LoS blockage process can be expressed as follows:
\begin{equation}\label{Intensity of Blockage}
{\lambda _B} = \lambda \mathbb{P}\left( {{H_B} \geqslant {H_{{\mathrm{LoS}}}}\left( l \right)} \right)\mathbb{P}\left( {\left| t \right| \leqslant {R_B}} \right),
\end{equation}
where $\left| t \right|$ denotes the distance between the center of the cylindrical obstacle's base and its projected position on the horizontal LoS path.
 
Since the effective blockage events are independent of each other, the expected number of obstacles that block the LoS path can be expressed as follows:
\begin{equation}\label{expectation_of_blockage_number}
\begin{split}
\mathbb{E}\left( \Lambda  \right) & = \iint_{\boldsymbol{\psi _B} \in \mathcal{A}} {{\lambda _B}d{\boldsymbol{\psi _B}}} \\
& = \lambda \int_0^D {\int_{ - \infty }^\infty  {\mathbb{P}\left( {{H_B} \geqslant {H_{{\mathrm{LoS}}}}\left( l \right)} \right)\mathbb{P}\left( {\left| t \right| \leqslant {R_B}} \right)dtdl} },
\end{split}
\end{equation}
where $\boldsymbol{\psi} _B$ denotes the position of the center of the cylindrical obstacle's base, and $\Lambda $ denotes the number of obstacles blocking the LoS path.

The LoS probability quantifies the likelihood that a direct, unblocked communication link exists between a PA and a UE. In the environments with randomly distributed obstacles, it serves as a key performance metric that captures the influence of blockage on overall communication reliability. According to the void probability property of PPP, the LoS probability corresponds to the probability that no obstacle intersects the line segment between the PA and the UE, which can be expressed as follows:
\begin{equation}\label{LoS_pr}
{\mathbb{P}_{{\mathrm{LoS}}}} = \mathbb{P}\left( {\Lambda  = 0} \right) = \exp \left( { - \mathbb{E}\left( \Lambda  \right)} \right).
\end{equation}

\section{On the Performance of PASS \\with LoS Blockage}\label{sec:Analysis}
In this section, the LoS blockage effect on the communication performance of PASS is analytically studied. The following three subsections present the LoS probabilities, outage probabilities, and ergodic rates under the proposed blockage models.

\subsection{LoS Probability}\label{subsec:LoS_Probability}
We introduce a RHRR blockage model, where the obstacle heights and radii are random variable. It is assumed that the obstacle heights follow a Gaussian distribution, i.e., ${H_B} \sim \mathcal{N}\left( \mu_B, \sigma_B^2 \right)$, where $\mu_B$ and $\sigma_B$ denote the mean and standard deviation, respectively. The obstacle radii are assumed to be uniformly distributed within a specified range, i.e., ${R_B} \sim \mathcal{U} \left( R_{B,\min}, R_{B,\max} \right)$, where $R_{B,\min}$ and $R_{B,\max}$ denote the minimum and maximum obstacle radii, respectively. Furthermore, obstacles are randomly located within the service area in accordance with a homogeneous PPP. The LoS probability of the RHRR blockage model can be characterized by the following lemma.
\begin{lemma}\label{lemma:LoS_pr_norm}
In the service area, given that obstacle locations follow a homogeneous PPP, we assume that obstacle heights follow a Gaussian distribution, and obstacle radii follow a uniform distribution, the LoS probability can be expressed as:
\begin{equation}\label{LoS_pr_norm}
{\mathbb{P}_{{\mathrm{LoS}}}} = \exp \left( { - \frac{{\left( {{R_{B,\min }} + {R_{B,\max }}} \right)\lambda DI}}{{H - {H_U}}}} \right),
\end{equation}
where $I = \int_{{H_U}}^H {\left[ {1 - \Phi \left( {\frac{{u - {\mu _B}}}{{{\sigma _B}}}} \right)} \right]du}$, and $\Phi(x)$ denotes the standard normal distribution function, defined as $\Phi \left( x \right) = \frac{1}{{\sqrt {2\pi } }}\int_{ - \infty }^x {{e^{ - \frac{{{k^2}}}{2}}}dk}$.
\begin{proof}
Please refer to Appendix A.
\end{proof}
\end{lemma}

While the RHRR model captures the full statistical features of the obstacle geometry, we introduce a DHDR blockage model to better highlight the impact of the number and location of obstacles, where both the obstacle radius and height are treated as deterministic values.\footnote{The RHRR and DHDR models are intended as tractable abstractions of practical blockage environments. On the one hand, the RHRR model reflects heterogeneous scenarios, where obstacles such as human bodies, furniture, or luggage may have different effective heights and lateral blocking ranges~\cite{7593259, 9247469, human_blockage1}. On the other hand, the DHDR blockage model represents a more homogeneous or controlled approximation, such as structural columns, storage tanks, or regularly arranged shelves~\cite{3gpp.38.901, 9492764}.} Specifically, as the standard deviation of obstacle heights approaches zero, all obstacles converge to a common height that satisfies $0 < H_B < H$. Likewise, if the obstacle radii are assumed to converge to their mean value, the LoS probability under the DHDR blockage model can be characterized by the following lemma.
\begin{lemma}\label{lemma:LoS_pr_constHB}
When the LoS blockage is mainly caused by random spatial distribution of obstacles in the service area, the LoS probability can be expressed as follows:
\begin{equation}\label{LoS_pr_constHB}
\mathbb{P}_{\mathrm{LoS}} = \exp \left( { - 2{R_B}\lambda D\frac{{{H_B} - {H_U}}}{{H - {H_U}}}} \right),\quad {H_U} < {H_B} < H.
\end{equation}
\begin{proof}
Please refer to Appendix B.
\end{proof}
\end{lemma}

\begin{remark}
Based on the results in~\eqref{LoS_pr_norm} and~\eqref{LoS_pr_constHB}, it is evident that the stochastic behavior of the performance analysis is attributed to key independent sources of randomness: the number, height, and radius of obstacles, as well as the locations of both the UE and obstacles.
\end{remark}

\subsection{Outage Probability}\label{subsec:OP}
The data rate of the UE can be expressed as follows:
\begin{equation}\label{Rate}
\begin{split}
\mathcal{R} & = {\log _2}\left( {1 + \frac{{P{{\left| {\tilde h} \right|}^2}}}{{{\sigma ^2}}}} \right) \\
     & = {\log _2}\left( {1 + \frac{{\kappa \gamma }}{{{\left\| {{\boldsymbol{\psi} ^{{\mathrm{PA}}}} - \boldsymbol{\psi} } \right\|}^2}}} \right),
\end{split}
\end{equation}
where $\sigma ^2$ denotes the noise power, $\eta = {\left( \frac{c}{4\pi {f_c}} \right)^2}$ and $\gamma  = \frac{\eta P}{{\sigma ^2}}$ are set for the sake of notational simplicity.

The outage probability of the UE indicates the likelihood of communication failure affected by potential LoS blockage and reflects the performance gain achieved by PASS, which can be expressed as follows:
\begin{equation}\label{outage}
\begin{split}
\mathbb{P}_{\mathrm{out}} & = \mathbb{P}\left( {\mathcal{R} \leqslant {\mathcal{R}_{{\mathrm{target}}}}} \right) \\
     & = \mathbb{P}\left( {{{\log }_2}\left( {1 + \frac{{\kappa \gamma }}{{{{\left\| \boldsymbol{\psi} ^{\mathrm{PA}} - \boldsymbol{\psi} \right\|}^2}}}} \right) \leqslant {\mathcal{R}_{{\mathrm{target}}}}} \right),
\end{split}
\end{equation}
where $\mathcal{R}_{\mathrm{target}}$ denotes the target data rate.
Since the LoS indicator $\kappa$ in~\eqref{effective_LoS_channel} represents whether the LoS path is blocked or not, the outage probability is formulated as follows:
\begin{equation}\label{outage_expand}
\begin{split}
\mathbb{P}_{\mathrm{out}} & = \mathbb{P}\left( \kappa  = 0 \right) + \mathbb{P}\left( {\frac{\gamma }{{{{\left\| {{\boldsymbol{\psi} ^{{\mathrm{PA}}}} - \boldsymbol{\psi} } \right\|}^2}}} \leqslant {2^{\mathcal{R}_{\mathrm{target}} - 1}},\kappa  = 1} \right) \\
     & = \mathbb{P}\left( {\kappa  = 0} \right) + \mathbb{P}\left( {\left\| {{\boldsymbol{\psi} ^{{\mathrm{PA}}}} - \boldsymbol{\psi} } \right\| \geqslant \varepsilon ,\kappa  = 1} \right),
\end{split}
\end{equation}
where $\varepsilon  = \sqrt {\tfrac{\gamma }{2^{\mathcal{R}_{\mathrm{target}}} - 1}}$.
By using the LoS probability derived in \textbf{Lemma~\ref{lemma:LoS_pr_norm}}, the outage probability is obtained as follows:
\begin{equation}\label{outage_substitute_PLoS}
\begin{split}
\mathbb{P}_{{\mathrm{out}}} & = \int_{\boldsymbol{\psi}  \in \mathcal{A}} {\left( {1 - {\mathbb{P}_{{\mathrm{LoS}}}}} \right)d\boldsymbol{\psi} }  + \int_{\boldsymbol{\psi}  \in \mathcal{A},\left\| {\boldsymbol{\psi} ^{{\mathrm{PA}}}} - \boldsymbol{\psi}  \right\| \geqslant \varepsilon } {{\mathbb{P}_{{\mathrm{LoS}}}}d\boldsymbol{\psi} } \\
     & = \int_{\boldsymbol{\psi}  \in \mathcal{A}} {\left( {1 - {e^{ - \beta D}}} \right)d\boldsymbol{\psi} }  + \int_{\boldsymbol{\psi}  \in \mathcal{A},\left\| {{\boldsymbol{\psi} ^{{\mathrm{PA}}}} - \boldsymbol{\psi} } \right\| \geqslant \varepsilon } {{e^{ - \beta D}}d\boldsymbol{\psi} },
\end{split}
\end{equation}
where $\beta  = \frac{\left( {R_{B,\min } + R_{B,\max }} \right)\lambda I}{{{H_d}}}$ denotes the exponential coefficient of the LoS probability, and ${H_d} = H - {H_U}$ denotes the difference in height between the PA and the UE.

To mitigate the free-space propagation loss, the PA is assumed to be activated on the waveguide at $\boldsymbol{\psi} ^{\mathrm{PA}} = [x,0,H]$, which is the position closest to the UE. By deriving the integrals in~\eqref{outage_substitute_PLoS}, the following theorem is obtained.
\begin{theorem}\label{theorem:OP_norm}
In the presence of LoS blockage arising from randomly distributed obstacles, whose radii and heights follow uniform and Gaussian distributions, respectively, the outage probability of the UE can be expressed as follows:
\begin{equation}\label{OP norm}
{\mathbb{P}_{{\mathrm{out}}}} = 1 + \frac{1}{{\beta W}}\left( {{e^{\beta {\tau _1}}} + {e^{ - \beta {\tau _2}}} - 2} \right),
\end{equation}
where ${\tau _1} = \max \left\{ { - \frac{W}{2}, - \sqrt {{\varepsilon ^2} - {H_d}^2} } \right\}$ and ${\tau _2} = \min \left\{ {\frac{W}{2},\sqrt {{\varepsilon ^2} - {H_d}^2} } \right\}$.
\begin{proof}
Please refer to Appendix C.
\end{proof}
\end{theorem}

The closed-form expression of the outage probability in~\eqref{OP norm} can be simplified under the condition of high transmit signal-to-noise ratio (SNR), which yields the following corollary.
\begin{corollary}\label{coro:OP norm asymptotic}
In the presence of LoS blockage arising from randomly distributed obstacles, whose radii and heights follow uniform and Gaussian distributions, respectively, there is a lower bound on the outage probability in the high-SNR regime, also as the asymptotic outage probability, which can be expressed as:
\begin{equation}\label{OP norm asymptotic}
{\mathbb{P}_{{\mathrm{out}}}} \approx 1 - \frac{2}{{\beta W}}\left( {1 - {e^{ - \frac{{\beta W}}{2}}}} \right).
\end{equation}
\begin{proof}
Under the condition of high transmit SNR, i.e., $\frac{P}{\sigma ^2} \to \infty$, when the target rate $\mathcal{R} _{\mathrm{target}}$ is a small and finite value, the equation $\sqrt {{\varepsilon ^2} - H_d^2}$ approaches $\varepsilon$ and is far greater than $\tfrac{W}{2}$, which leads to $\tau_1 = - \tfrac{W}{2}$ and $\tau_2 = \tfrac{W}{2}$. Moreover, since the outage probability in~\eqref{OP norm} involves the exponential terms $e^{\beta \tau_1}$ and $e^{-\beta \tau_2}$, which are both monotonic functions, the outage probability reaches its lower bound when $\tau_1$ and $\tau_2$ take their boundary values, i.e., $\tau_1 = - \tfrac{W}{2}$ and $\tau_2 = \tfrac{W}{2}$. Hence, the proof is complete.
\end{proof}
\end{corollary}

\begin{remark}\label{remark:OP norm}
According to the results in~\eqref{OP norm} and its asymptotic results in~\eqref{OP norm asymptotic}, the outage probability is positively correlated not only with the spatial distribution and geometric dimensions of obstacles, but also with the width of the service area.
\end{remark}

\begin{proposition}\label{propo:OP W approches infty}
In the high-SNR regime, when the width of the service area is infinitely large, the outage probability of the UE can be approximated as follows:
\begin{equation}\label{OP W approches infty}
\mathop {\lim }\limits_{W \to \infty } {\mathbb{P}_{{\mathrm{out}}}} \approx 1 - \mathop {\lim }\limits_{W \to \infty } \frac{2}{{\beta W}}\left( {1 - {e^{ - \frac{{\beta W}}{2}}}} \right) = 1.
\end{equation}
\begin{proof}
We set $k = \frac{{\beta W}}{2}$. For $k > 0$, the inequality $0 < \frac{{1 - {e^{ - k}}}}{k} < \frac{1}{k}$ holds. When the width of the service area satisfies $W \to \infty$, then $\frac{1}{k} \to 0$. According to the Squeeze Theorem~\cite{sohrab2003basic}, the equation $\mathop {\lim }\limits_{W \to \infty } \frac{{1 - {e^{ - k}}}}{k} = 0$ holds true, and the proof is complete.
\end{proof}
\end{proposition}

\begin{remark}\label{remark:OP norm When W infty}
An interesting observation from~\eqref{OP W approches infty} is that, under a fixed obstacle intensity, the outage probability increases with the width of the service area and eventually approaches one, which highlights the dominant impact of spatially distributed obstacles on the LoS availability of PASS in wide-area deployments.
\end{remark}

To further illustrate the impact of spatial intensity characteristics of obstacles, we focus on a special case of \textbf{Theorem~\ref{theorem:OP_norm}}, where all obstacles share identical radii and heights. In this case, the outage probability of the UE admits a simplified closed-form expression, as stated in the following theorem.
\begin{theorem}\label{theorem:OP const}
In the presence of LoS blockage arising from randomly distributed obstacles with identical radii and identical heights, the outage probability of the UE can be expressed as follows:
\begin{equation}\label{OP const}
{\mathbb{P}_{\mathrm{out}}} = 1 + \frac{1}{{\rho W}}\left( {e^{\rho {\tau _1}} + {e^{ - \rho {\tau _2}}} - 2} \right),
\end{equation}
where $\rho  = 2{R_B}\lambda \frac{H_B - H_U}{H_d}$.
\begin{proof}
The results in~\eqref{OP const} are direct simplification of \textbf{Theorem~\ref{theorem:OP_norm}} by substituting the LoS probability given in \textbf{Lemma~\ref{lemma:LoS_pr_constHB}}. Hence, the proof is complete.
\end{proof}
\end{theorem}

\begin{corollary}\label{coro:OP const asymptotic}
In the presence of LoS blockage arising from randomly distributed obstacles with identical radii and identical heights, there is a lower bound on the outage probability in the high-SNR regime, also as the asymptotic outage probability, which can be expressed as:
\begin{equation}\label{OP const asymptotic}
{\mathbb{P}_{\mathrm{out}}} \approx 1 - \frac{2}{{\rho W}}\left( {1 - {e^{ - \frac{{\rho W}}{2}}}} \right).
\end{equation}
\begin{proof}
Similar to \textbf{Corollary~\ref{coro:OP norm asymptotic}}, the results in~\eqref{OP const asymptotic} are readily derived.
\end{proof}
\end{corollary}

\begin{proposition}\label{propo:diversity order}
From \textbf{Corollary~\ref{coro:OP norm asymptotic}} and \textbf{Corollary~\ref{coro:OP const asymptotic}}, the diversity order for the scenario with LoS blockage can be obtained by using the high-SNR approximation, and the diversity order of the UE is given by
\begin{equation}\label{diversity order}
w =  - \mathop {\lim }\limits_{\frac{P}{{\sigma ^2}} \to \infty } \frac{{\log {\mathbb{P}_{{\mathrm{out}}}}}}{{\log \frac{P}{{\sigma ^2}}}} \approx 0.
\end{equation}
\begin{proof}
According to~\eqref{OP norm asymptotic} and~\eqref{OP const asymptotic}, it is clear that the outage probability is a constant value between 0 and 1 for a fixed obstacle intensity $\lambda$ and a fixed area width $W$ in the high-SNR regime, independent of the transmit SNR. Hence, the proof is complete.
\end{proof}
\end{proposition}

\begin{remark}\label{remark:diversity order}
The diversity order in~\eqref{diversity order} indicates that, in the presence of LoS blockage caused by randomly spatial distributed obstacles in the service area, the transmit SNR has almost no effect on the outage probability in the high-SNR regime.
\end{remark}

In contrast to the PA that can be activated transversely along the waveguide, the conventional fixed-position antenna is assumed to be located at ${\boldsymbol{\psi} ^{\mathrm{conv}}} = \left[ {0,0,H} \right]$, i.e., over the geometric center of the service area, chosen under the assumption that the UE is randomly and uniformly distributed within the service area, thereby providing the most equitable and representative basis for performance comparison. The outage probability of the UE achieved by CASS can be expressed as follows:
\begin{equation}\label{outage_conv}
\mathbb{P}_{\mathrm{out}}^{\mathrm{conv}} = \mathbb{P}\left( {{\log }_2}\left( {1 + \frac{{\kappa \gamma }}{{{{\left\| {{\boldsymbol{\psi} ^{{\mathrm{conv}}}} - \boldsymbol{\psi} } \right\|}^2}}}} \right) \leqslant {\mathcal{R}_{{\mathrm{target}}}} \right).
\end{equation}

Under the condition of high transmit SNR, similar to the mathematical treatment in \textbf{Corollary~\ref{coro:OP norm asymptotic}}, the outage probability of the UE achieved by CASS can be approximated as follows:
\begin{equation}\label{outage_conv_approximate}
\mathbb{P}_{{\mathrm{out}}}^{{\mathrm{conv}}} \approx \frac{1}{{LW}}\int_{ - \frac{W}{2}}^{\frac{W}{2}} {\int_{ - \frac{L}{2}}^{\frac{L}{2}} {\left( {1 - {e^{ - \beta \sqrt {{x^2} + {y^2}} }}} \right)dxdy} }.
\end{equation}

Comparing the outage probability affected by LoS blockage between PASS and CASS in the high-SNR regime, the following proposition can be derived.
\begin{proposition}\label{propo:OP difference}
For the scenario involving the random LoS blockage, PASS achieves an outage probability that is strictly lower than that of CASS in the high-SNR regime.
\begin{proof}
The outage probability achieved by PASS over that of CASS can be expressed as follows: 
\begin{equation}\label{outage_diff}
\begin{aligned}
& \mathbb{P}_{\mathrm{out}}^{\mathrm{conv}} - \mathbb{P}_{\mathrm{out}} \\
& \approx \frac{2}{W}\int_0^{\frac{W}{2}} {\left( {e^{ - \beta \sqrt {y^2} } - \frac{2}{L}\int_0^{\frac{L}{2}} {{e^{ - \beta \sqrt {x^2 + y^2} }}dx} } \right)dy}.
\end{aligned}
\end{equation}

Note that for $\beta > 0$ and a constant $a$ satisfying $a > 0$, the following inequality holds as follows:
\begin{equation}\label{outage_diff_proof}
\frac{2}{L}\int_0^{\frac{L}{2}} {{e^{ - \beta \sqrt {x^2 + a} }}dx}  < \frac{2}{L}\int_0^{\frac{L}{2}} {{e^{ - \beta \sqrt a }}dx}  = {e^{ - \beta \sqrt a }},
\end{equation}
which means $\mathbb{P}_{\mathrm{out}}^{\mathrm{conv}} - {\mathbb{P}_{\mathrm{out}}} > 0$, and the proof is complete.
\end{proof}
\end{proposition}

\subsection{Ergodic Rate}\label{subsec:ER}

The ergodic rate of the UE can be expressed as follows:
\begin{equation}\label{ER}
\begin{split}
\bar {\mathcal{R}} & = \mathbb{E}\left( {\mathbb{P}\left( {\kappa  = 1} \right){{\log }_2}\left( {1 + \frac{\kappa \gamma }{{{{\left\| {{\boldsymbol{\psi} ^{{\mathrm{PA}}}} - \boldsymbol{\psi} } \right\|}^2}}}} \right)} \right) \\
     & = \mathbb{E}\left( {{\mathbb{P}_{{\mathrm{LoS}}}}{{\log }_2}\left( {1 + \frac{\gamma }{{{{\left\| {{\boldsymbol{\psi} ^{{\mathrm{PA}}}} - \boldsymbol{\psi} } \right\|}^2}}}} \right)} \right).
\end{split} 
\end{equation}

Assuming that the PA is activated at ${\boldsymbol{\psi} ^{{\mathrm{PA}}}} = [x,0,H]$, and the UE is randomly located with the service area, the ergodic rate performance can be further investigated, which derives the following theorem.

\begin{theorem}\label{theorem:ER norm}
In the presence of LoS blockage arising from randomly distributed obstacles, whose radii and heights follow uniform and Gaussian distributions, respectively, the ergodic rate of the UE can be expressed as follows:
\begin{equation}\label{ER norm}
\bar{\mathcal{R}} = \frac{1}{W}\int_{ - \frac{W}{2}}^{\frac{W}{2}} {{e^{ - \beta \left| y \right|}}{{\log }_2}\left( {1 + \frac{\gamma }{y^2 + H_d^2}} \right)dy}.
\end{equation}
\begin{proof}
Based on \textbf{Lemma~\ref{lemma:LoS_pr_norm}} and the assumption that the UE is uniformly distributed in the service area, the expectation in~\eqref{ER} can be transformed into a definite integral, and the proof is complete.
\end{proof}
\end{theorem}

Since obtaining a closed-form expression of the ergodic rate in~\eqref{ER norm} is challenging, we use a Taylor expansion~\cite{Table_of_integrals} in the following corollary.
\begin{corollary}\label{coro:ER norm approxi}
In the presence of LoS blockage arising from the random spatial distribution of obstacles, whose radii and heights follow uniform and Gaussian distributions, respectively, the ergodic rate of the UE can be approximated as follows:
\begin{equation}\label{ER norm approxi}
\begin{split}
\bar {\mathcal{R}} \approx & \frac{2}{{\beta W}}\left( {1 - {e^{ - \frac{{\beta W}}{2}}}} \right){\log _2}\left( {1 + \frac{\gamma }{\zeta }} \right) \\
     & - \frac{{\gamma \left( {\frac{1}{\beta }\left( {1 - {e^{ - \frac{{\beta W}}{2}}}} \right) - W\left( {1 + {e^{ - \frac{{\beta W}}{2}}}} \right)} \right)}}{{\beta \zeta \left( {\gamma  + \zeta } \right)\ln 2}},
\end{split}
\end{equation}
where $\zeta  = H_d^2 + \frac{W^2}{16}$.
\begin{proof}
Please refer to Appendix D.
\end{proof}
\end{corollary}

To gain a deep insight into the trend of rate performance scaling, the high-SNR slope of the ergodic rate in the high-SNR regime is given in the following proposition.
\begin{proposition}\label{propo:high_SNR slope}
For the scenario involving the random LoS blockage, when the transmit SNR is sufficiently high, the high-SNR slope is characterized as follows:
\begin{equation}\label{high-SNR slope}
\mathcal{S} = \frac{2}{\beta W}\left( {1 - {e^{ - \frac{{\beta W}}{2}}}} \right).
\end{equation}
\begin{proof}
Please refer to Appendix E.
\end{proof}
\end{proposition}

To further illustrate the impact of spatial intensity characteristics of obstacles, we consider a special case of \textbf{Theorem~\ref{theorem:ER norm}}, where all obstacles share identical radii and heights. In this case, the ergodic rate of the UE admits a simplified closed-form expression, as stated in the following theorem.
\begin{theorem}\label{theorem:ER const}
In the presence of LoS blockage arising from randomly distributed obstacles with identical radii and identical heights, the ergodic rate of the UE can be expressed as follows:
\begin{equation}\label{ER const}
\bar{\mathcal{R}} = \frac{2}{W}\int_0^{\frac{W}{2}} {{e^{ - \rho y}}{{\log }_2}\left( {1 + \frac{\gamma }{y^2 + H_d^2}} \right)dy}.
\end{equation}
\begin{proof}
The results in~\eqref{ER const} are direct simplification of \textbf{Theorem~\ref{theorem:ER norm}} by substituting the LoS probability in \textbf{Lemma~\ref{lemma:LoS_pr_constHB}}. Hence, the proof is complete.
\end{proof}
\end{theorem}

\begin{corollary}\label{coro:ER const approxi}
In the presence of LoS blockage arising from randomly distributed obstacles with identical radii and identical heights, the ergodic rate of the UE can be approximated as follows:
\begin{equation}\label{ER const approxi}
\begin{split}
\bar{\mathcal{R}} \approx & \frac{2}{{\rho W}}\left( {1 - {e^{ - \frac{{\rho W}}{2}}}} \right){\log _2}\left( {1 + \frac{\gamma }{\zeta }} \right) \\
     & - \frac{{\gamma \left( {\frac{1}{\rho }\left( {1 - {e^{ - \frac{{\rho W}}{2}}}} \right) - W\left( {1 + {e^{ - \frac{{\rho W}}{2}}}} \right)} \right)}}{{\rho \zeta \left( {\gamma  + \zeta } \right)\ln 2}}.
\end{split}
\end{equation}
\begin{proof}
Similar to \textbf{Corollary~\ref{coro:ER norm approxi}}, the results in~\eqref{ER const approxi} are readily derived.
\end{proof}
\end{corollary}

\begin{remark}\label{remark:Ergodic Rate when HB is constant}
According to the results in~\eqref{ER norm} and~\eqref{ER const}, the ergodic rate is negatively correlated with both the spatial density of obstacles and the service area width, and it also depends on the transmit SNR. Furthermore, based on \textbf{Proposition~\ref{propo:high_SNR slope}}, in an area of finite width with a finite density of obstacles, the ergodic rate is shown to increase linearly in the high-SNR regime.
\end{remark}

The ergodic rate of the UE achieved by CASS can be expressed as follows:
\begin{equation}\label{Rate_Ergodic_conv}
\begin{split}
{\bar {\mathcal{R}}^{{\mathrm{conv}}}} & = \mathbb{E}\left( {\mathbb{P}\left( {\kappa  = 1} \right){{\log }_2}\left( {1 + \frac{{\kappa \gamma }}{{{{\left\| {{\boldsymbol{\psi} ^{{\mathrm{conv}}}} - \boldsymbol{\psi} } \right\|}^2}}}} \right)} \right) \\
     & = \mathbb{E}\left( {{\mathbb{P}_{{\mathrm{LoS}}}}{{\log }_2}\left( {1 + \frac{\gamma }{{{{\left\| {{\boldsymbol{\psi} ^{{\mathrm{conv}}}} - \boldsymbol{\psi} } \right\|}^2}}}} \right)} \right).
\end{split}
\end{equation}

Similar to \textbf{Theorem~\ref{theorem:ER norm}}, by applying the random process of blockages and the uniform distribution assumption of the UE in the service area, the ergodic rate achieved by CASS can be further written as follows:
\begin{equation}\label{Rate_Ergodic_conv_2}
\begin{split}
{\bar {\mathcal{R}}}^{\mathrm{conv}} = & \frac{1}{{LW}}\int_{ - \frac{W}{2}}^{\frac{W}{2}} {\int_{ - \frac{L}{2}}^{\frac{L}{2}} {e^{ - \beta \sqrt {x^2 + y^2} }} } \\
     & \times {\log _2}\left( {1 + \frac{\gamma }{x^2 + y^2 + H_d^2}} \right)dxdy.
\end{split}
\end{equation}

Comparing the ergodic rate affected by LoS blockage between PASS and CASS, the following proposition can be derived.
\begin{proposition}\label{propo:Ergodic Rate difference}
For the scenario involving the random LoS blockage, PASS achieves an ergodic rate that is strictly greater than that of CASS.
\begin{proof}
The ergodic rate achieved by PASS over that of conventional antennas is written as follows:
\begin{equation}\label{Ergodic Rate_diff}
\begin{gathered}
  \bar{\mathcal{R}} - {{\bar{\mathcal{R}}}^{{\mathrm{conv}}}} = \frac{2}{W}\int_0^{\frac{W}{2}} {\left[ {{e^{ - \beta y}}{{\log }_2}\left( {1 + \frac{\gamma }{y^2 + H_d^2}} \right)} \right.}  \hfill \\
   - \frac{2}{L}\left. {\int_0^{\frac{L}{2}} {{e^{ - \beta \sqrt {{x^2} + {y^2}} }}{{\log }_2}\left( {1 + \frac{\gamma }{x^2 + y^2 + H_d^2}} \right)} dx} \right]dy \hfill. \\ 
\end{gathered}
\end{equation}

For $\beta > 0$, $\gamma > 0$ and a constant $a$ satisfying $a > 0$, we set ${g_3}\left( x \right) = {e^{ - \beta \sqrt {{x^2} + a} }}{\log _2}\left( {1 + \frac{\gamma }{{{x^2} + a}}} \right)$. Note that ${g_3}\left( x \right)$ is a monotonically decreasing function when $x > 0$, the following inequality holds:
\begin{equation}\label{Erogdic Rate_diff_proof}
\frac{2}{L}\int_0^{\frac{L}{2}} {{g_3}\left( x \right)} dx < \frac{2}{L}\int_0^{\frac{L}{2}} {{g_3}\left( 0 \right)} dx = {e^{ - \beta \sqrt a }}{\log _2}\left( {1 + \frac{\gamma }{a}} \right),
\end{equation}
which leads to $\bar {\mathcal{R}} - {{\bar {\mathcal{R}}}^{{\mathrm{conv}}}} > 0$. This completes the proof.
\end{proof}
\end{proposition}

\begin{remark}\label{remark:Performance gain diff}
According to \textbf{Proposition~\ref{propo:OP difference}} and \textbf{Proposition~\ref{propo:Ergodic Rate difference}}, it becomes evident that the fundamental mechanism behind performance improvement is the dynamic repositioning capability of PASS.
\end{remark}

\section{Numerical Results}\label{sec:Sim}

\begin{table}[t!]
  \begin{center}
  \caption{Simulation Parameters}\footnotesize\label{tab:simPara}
  \begin{tabular}{|l|l|l|}    
    \hline
    \textbf{Parameter} & \textbf{Description} & \textbf{Value} \\
    \hline
    $f_c$ & Carrier frequency & $28$ GHz \\
    \hline
    $\mathcal{R}_{\mathrm{target}}$ & Target data rate & $1$ bit/s/Hz \\    
    \hline
    $BW$ & Bandwidth & $10$ MHz \\
    \hline
    $P$ & Transmit power & $0$ dBm \\
    \hline
    $n_{\mathrm{eff}}$ & Effective refractive index of the waveguide & $1.4$ \\
    \hline
    $H$ & Height of the waveguide & $3$ m \\    
    \hline
    $L$ & Length of the service area & $10$ m \\
    \hline
    $W$ & Width of the service area & $10$ m \\
    \hline
    $H_U$ & Height of the UE & $0.2$ m \\
    \hline
    $\lambda$ & Intensity of obstacles & $0.1$ m$^{ - 2}$ \\
    \hline
    $\mu _B$ & Mean height of obstacles & $1.6$ m \\
    \hline
    $\sigma _B$ & Standard deviation of obstacle heights & $0.2$ m \\
    \hline
    $R_{B,\min}$ & Minimum obstacle radius & $0.1$ m \\
    \hline
    $R_{B,\max}$ & Maximum obstacle radius & $0.3$ m \\
    \hline
  \end{tabular}
  \end{center}
\end{table}

This section provides numerical evaluations of the performance of PASS with LoS blockage, where analytical results are validated against Monte Carlo simulations. The AWGN power is configured as $\sigma ^2 = - 174 + 10{\log _{10}}\left( BW \right)$ dBm. Unless otherwise specified, the parameters are used in Table~\ref{tab:simPara}~\cite{ding2025flexible}.
Except Fig.~\ref{fig:OP3_with_lambda_compBlockModel} and Fig.~\ref{fig:ER3_with_lambda_compBlockModels}, the RHRR blockage model is used by default. In addition, the performance of CASS in equivalent conditions are also provided for comparison.

\subsection{Outage Probability}

\begin{figure}[t!]
  \centering
  \includegraphics[width= 3.1 in]{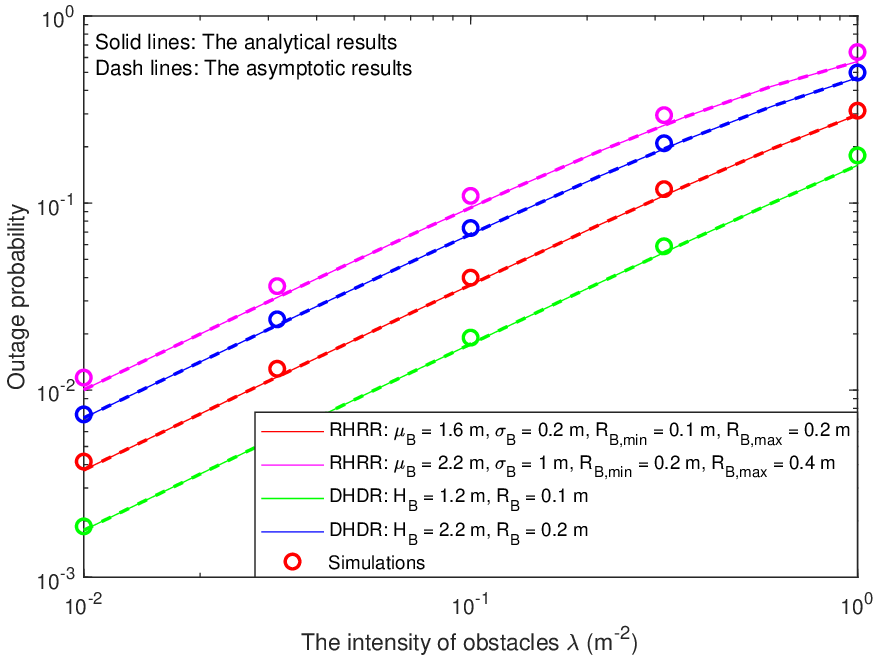}
  \caption{Outage probability versus obstacle intensity under different blockage model settings in the RHRR and DHDR models, respectively. The analytical and asymptotic results for the RHRR models are derived from~\eqref{OP norm} and~\eqref{OP norm asymptotic}, respectively. The analytical and asymptotic results for the DHDR models are derived from~\eqref{OP const} and~\eqref{OP const asymptotic}, respectively.}
  \label{fig:OP3_with_lambda_compBlockModel}
\end{figure}

\emph{1) Outage probability under different blockage models:} Fig.~\ref{fig:OP3_with_lambda_compBlockModel} illustrates the outage probability performance under different blockage models, where both the obstacle height and radius follow either statistical distributions or deterministic values. The analytical results match the Monte Carlo simulations well across the entire obstacle intensity range, thereby validating the correctness of the proposed RHRR and DHDR blockage models, and verifying the analysis in \textbf{Theorem~\ref{theorem:OP_norm}} and \textbf{Theorem~\ref{theorem:OP const}}. Specifically, the analytical results perfectly overlaps with their corresponding asymptotic results in the high-SNR regime, which demonstrates the tightness and accuracy of \textbf{Corollary~\ref{coro:OP norm asymptotic}} and \textbf{Corollary~\ref{coro:OP const asymptotic}}. Therefore, the comparisons of the asymptotic and analytical results in the high-SNR regime are omitted in the subsequent simulations. It can also be observed that the outage probability consistently increases with the intensity of obstacles, which reflects the fact that denser obstacle environments make the LoS path more likely to be blocked. Moreover, for a fixed obstacle intensity, scenarios with taller obstacles or larger obstacle radii yield higher outage probabilities, which highlights the significant role of obstacle geometry in LoS blockage effect.

\begin{figure}[t!]
  \centering
  \includegraphics[width= 3.1 in]{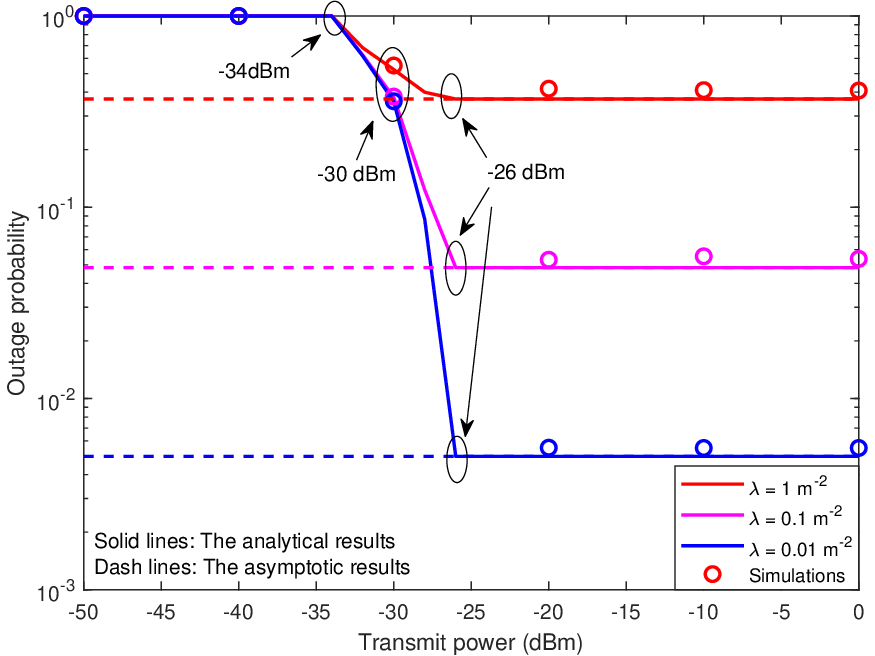}
  \caption{Outage probability versus transmit power under different obstacle intensities.}
  \label{fig:OP3_with_TPOW_compLAMBDA}
\end{figure}

\emph{2) Outage probability with different transmit power:} Fig.~\ref{fig:OP3_with_TPOW_compLAMBDA} illustrates the outage probability as a function of the transmit power for three different levels of obstacle intensities ($\lambda = 1, 0.1, 0.01$ m$^{ - 2}$). The analytical results closely match the Monte Carlo simulations, and the asymptotic results effectively characterize the lower performance bound of PASS. When the transmit power is sufficiently high, since random processes of LoS blockage are determined primarily by geometric relations between the obstacles, PA, and the UE, the outage probability remains nearly constant, which confirms our \textbf{Proposition~\ref{propo:diversity order}}. For a given transmit power, a higher obstacle intensity yields a larger outage probability, as a densely obstructed environment increases the probability of random LoS blockage. In contrast, in the region of low transmit power (i.e., below -26 dBm), since the received SNR of the UE is too weak to support the target rate, the outage probability increases sharply as the transmit power decreases. Then, complete outage occurs below -34 dBm. Overall, the impact of the transmit SNR on the outage probability is limited. In the low-SNR regime, it is necessary to increase transmit power to sustain the target data rate and avoid complete outage. However, once the received SNR of the UE exceeds the threshold required for reliable communication, further increasing the transmit power has almost no effect on the outage probability. From a physical perspective, the outage probability is dominated by the spatial intensity and geometry of obstacles in the high-SNR regime, which verifies our \textbf{Remark~\ref{remark:diversity order}}.

\begin{figure}[t!]
  \centering
  \includegraphics[width= 3.1 in]{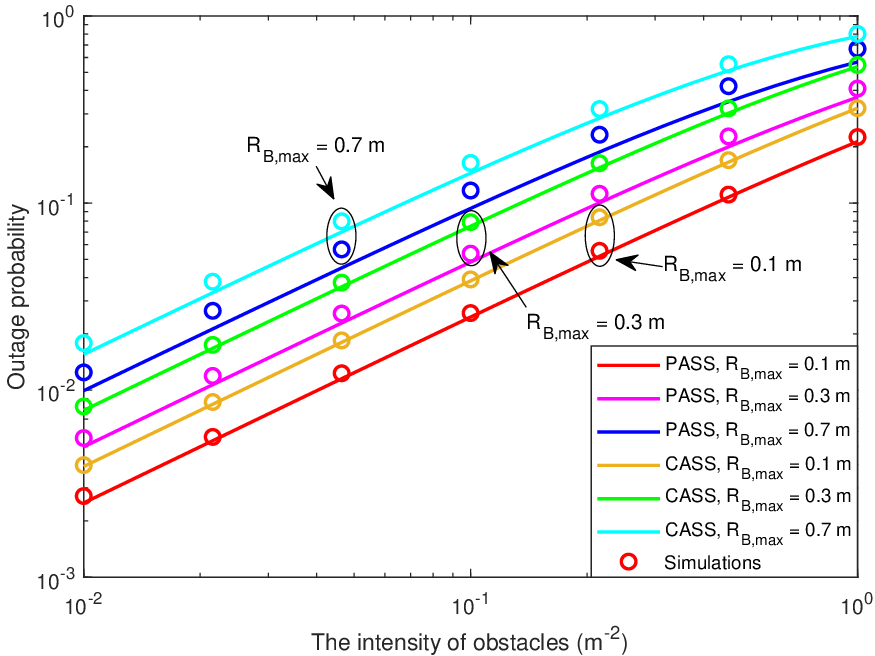}
  \caption{Comparison of the outage probability for PASS and CASS as a function of the obstacle intensity under different maximum obstacle radii.}
  \label{fig:OP3_with_Lambda_compRB}
\end{figure}

\emph{3) Outage probability affected by the density and radius of obstacles:} Fig.~\ref{fig:OP3_with_Lambda_compRB} illustrates the outage probability as a function of the obstacle intensity under various maximum obstacle radii ($R_{B,\max}=0.1,0.3,0.7$ m), which compares PASS with CASS. The obstacle intensity and radius respectively determine the number of obstacles and the area of individual obstacles, thus characterize the effect of the obstacle area ratio within the service area on the outage probability. Across all settings, the outage probability increases monotonically with obstacle intensity, confirming the fact that densely obstructed environments contain a large number of potential blockers within a fixed service area, thereby increasing the probability of LoS blockage. Similarly, for a given obstacle intensity, larger maximum radii correspond to ``fatter'' obstacles that occupy a greater fraction of the service area, further raising the risk of LoS blockage. Furthermore, it can be observed that PASS outperforms CASS, owing to its advantage of flexibly shifting antenna positions in a large scale. Therefore, the analytical results are generally accurate, which demonstrates the accuracy and reliability of the proposed blockage models under moderate conditions.

\begin{figure}[t!]
  \centering
  \includegraphics[width = 3.1 in]{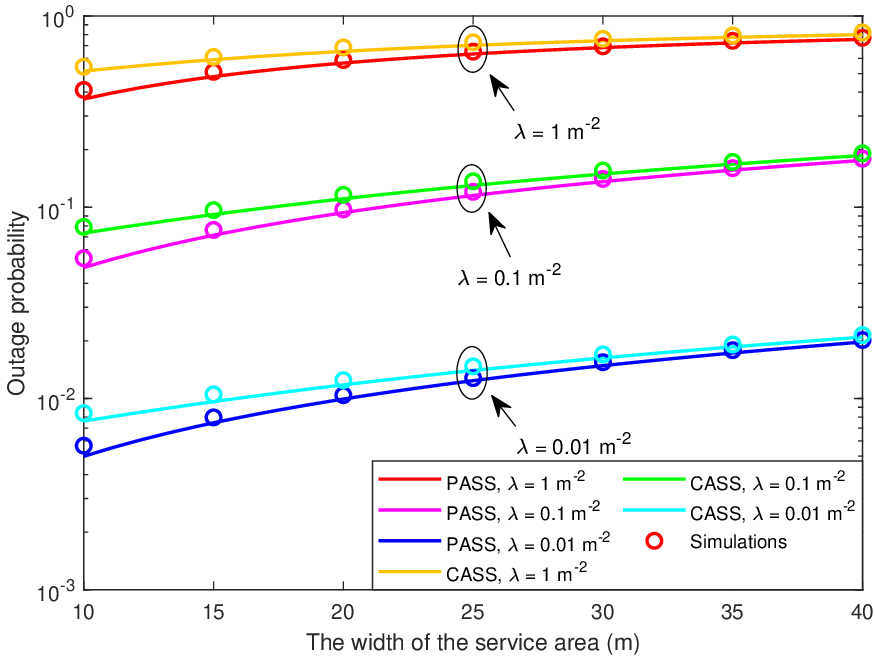}
  \caption{Comparison of the outage probability for PASS and CASS as a function of the width of the service area under different obstacle intensities.}
  \label{fig:OP_with_W_compLAMBDA}
\end{figure}

\emph{4) Outage probability with different area widths:} Fig.~\ref{fig:OP_with_W_compLAMBDA} illustrates the outage probability as a function of area width under various obstacle intensities ($\lambda = 1, 0.1, 0.01$ m$^{ - 2}$), which compares PASS with CASS. In all configurations, the outage probability increases with either a wider service area or a higher obstacle intensity. This is because enlarging the service area or increasing obstacle intensity inevitably introduces more obstacles within the coverage region, which raises the probability of LoS blockage effect. For instance, in a densely obstructed environment where the intensity of obstacles reaches 1 m$^{ - 2}$, the outage probability exceeds $10^{ - 1}$, regardless of the area width, which indicates that the impact of LoS blockage can no longer be ignored. Specifically, in practical large-scale environments, such as open squares and indoor halls, the extended propagation distance exposes the LoS path to a greater number of spatially distributed obstacles, resulting in a higher probability of blockage, which is consistent with the analytical insight in \textbf{Remark~\ref{remark:OP norm When W infty}}. Moreover, as the area width increases, the difference in outage probability between PASS and CASS gradually diminishes due to the intensification of large-scale fading. Nevertheless, it is evident that PASS performs better than CASS, even in wide-area deployments of PAs with LoS blockage.

\subsection{Ergodic Rate}

\begin{figure}[t!]
  \centering
  \includegraphics[width= 3.1 in]{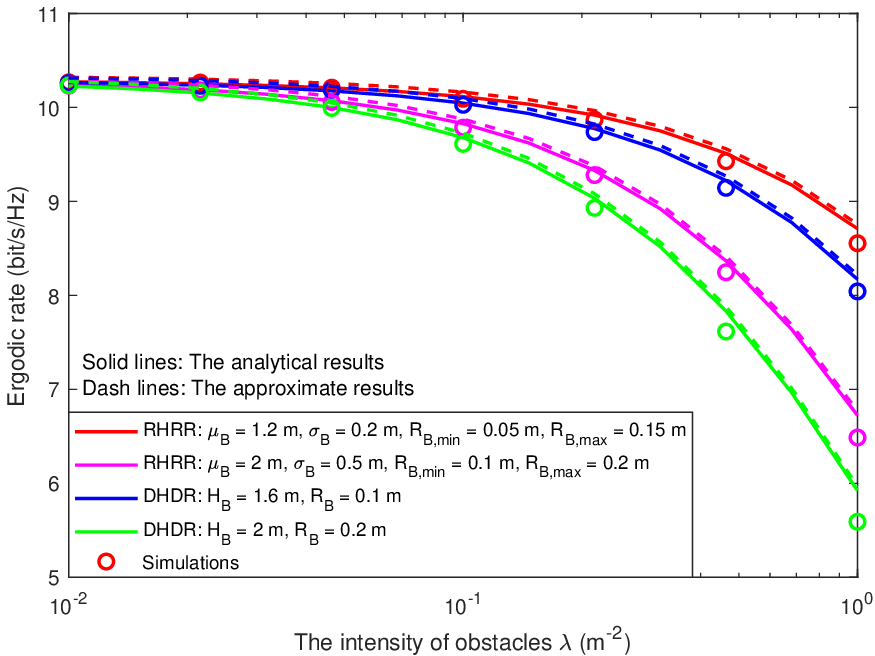}
  \caption{Ergodic rate versus obstacle intensity under different blockage model settings in the RHRR and DHDR models, respectively. The analytical and approximate results for the RHRR models are derived from~\eqref{ER norm} and~\eqref{ER norm approxi}, respectively. The analytical and approximate results for the DHDR models are derived from~\eqref{ER const} and~\eqref{ER const approxi}, respectively.}
  \label{fig:ER3_with_lambda_compBlockModels}
\end{figure}

\emph{5) Ergodic rate under different blockage models:} Fig.~\ref{fig:ER3_with_lambda_compBlockModels} illustrates the ergodic rate as a function of the obstacle intensity under different blockage models, where both the obstacle height and radius follow either statistical distributions or deterministic values. The analytical results and Monte Carlo simulations exhibit consistent performance trends, which supports the effectiveness of the expressions derived in \textbf{Theorem~\ref{theorem:ER norm}} and \textbf{Theorem~\ref{theorem:ER const}}. In particular, the ergodic rate decreases monotonically with the obstacle intensity under each parameter setting, which indicates that denser obstacles lead to more severe LoS blockage and hence degrade the achievable rate. Furthermore, the approximate results are close to their corresponding analytical results in the high-SNR regime, which demonstrates the accuracy and practical applicability of \textbf{Corollary~\ref{coro:ER norm approxi}} and \textbf{Corollary~\ref{coro:ER const approxi}}. It can also be seen that a higher obstacle height and radius result in a lower ergodic rate for the same level of obstacle intensity, which confirms that obstacle geometry has an impact on the performance degradation caused by LoS blockage.

\begin{figure}[t!]
  \centering
  \includegraphics[width= 3.1 in]{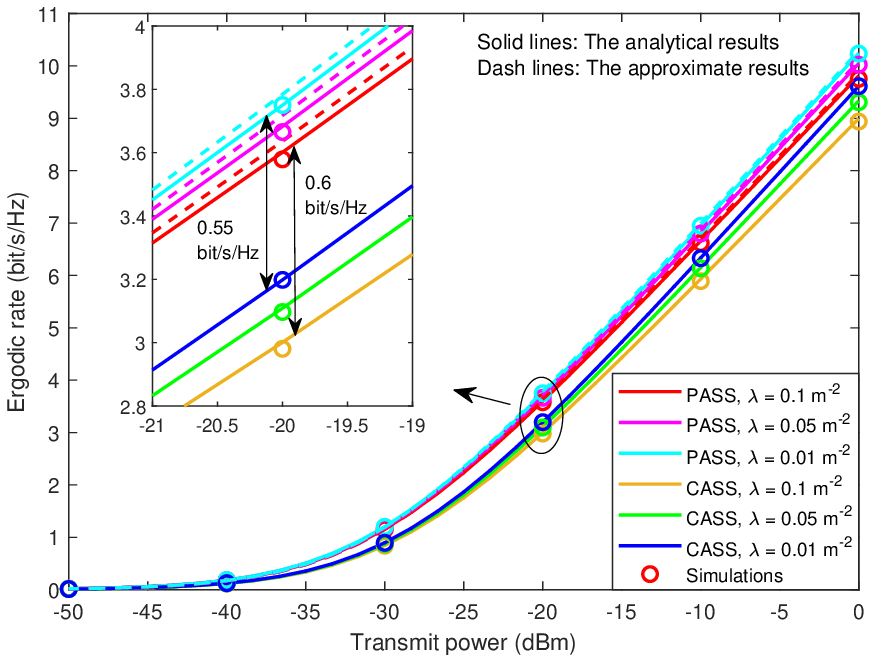}
  \caption{Comparison of the ergodic rate for PASS and CASS as a function of the transmit power under different obstacle intensities.}
  \label{fig:ER3_with_TPOW_compLAMBDA}
\end{figure}

\emph{6) Ergodic rate with different transmit power:} Fig.~\ref{fig:ER3_with_TPOW_compLAMBDA} illustrates the ergodic rate as a function of transmit power for three different levels of obstacle intensities ($\lambda = 0.1,0.05,0.01$ m$^{ - 2}$), comparing PASS with CASS. For each parameter setting, the ergodic rate grows steadily as the transmit power rises, which signifies the anticipated SNR-driven growth. Under the high transmit power condition (approximately $-$20 dBm and above), the ergodic rate exhibits a nearly linear growth trend, with an approximate slope across all settings, which confirms \textbf{Proposition~\ref{propo:high_SNR slope}}. Furthermore, PASS consistently outperforms CASS over the entire transmit power range and for all levels of obstacle intensity. For instance, at $-$20 dBm, the ergodic rate differences between PASS and CASS are approximately 0.55 bit/s/Hz and 0.6 bit/s/Hz, respectively, with obstacle intensities of 0.01 m$^{ - 2}$ and 0.1 m$^{ - 2}$, respectively. Additionally, it is clear that increasing the obstacle intensity results in a decrease in the ergodic rate, which confirms our \textbf{Remark~\ref{remark:Ergodic Rate when HB is constant}}.

\begin{figure}[t!]
  \centering
  \includegraphics[width= 3.1 in]{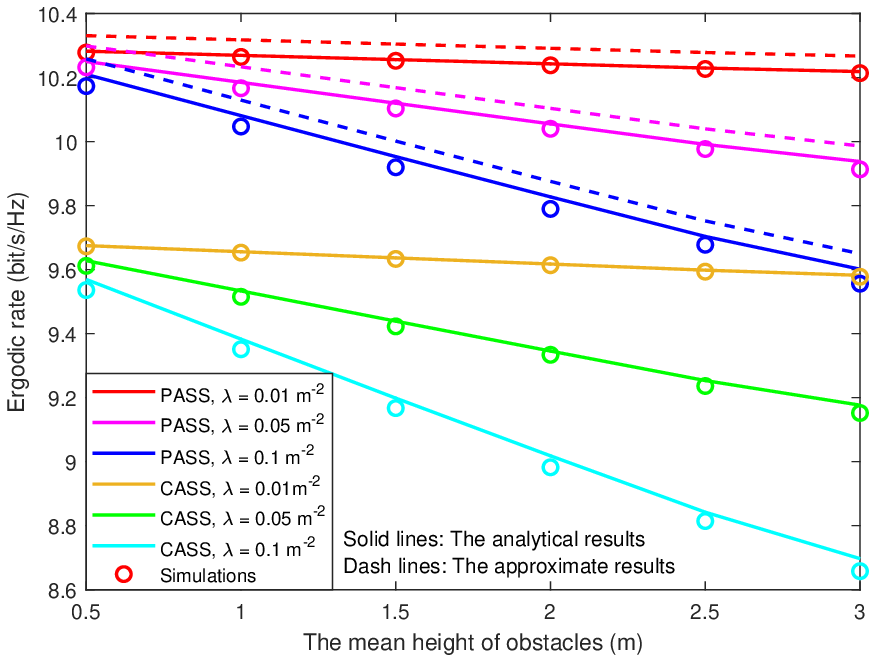}
  \caption{Comparison of the ergodic rate for PASS and CASS as a function of the mean obstacle height under different obstacle intensities.}
  \label{fig:ER3_with_HBmean_compHBstd}
\end{figure}

\emph{7) Ergodic rate affected by different obstacle heights:} Fig.~\ref{fig:ER3_with_HBmean_compHBstd} illustrates the ergodic rate as a function of the mean obstacle height for three different levels of obstacle intensities ($\lambda = 0.1,0.05,0.01$ m$^{ - 2}$), which compares PASS with CASS. Across all settings, the ergodic rate decreases monotonically as the mean obstacle height increases, which reflects that taller obstacles tend to block the LoS path. For a fixed obstacle intensity, PASS consistently outperforms the CASS in ergodic rate, which confirms our \textbf{Proposition~\ref{propo:Ergodic Rate difference}}. Moreover, as the obstacle intensity increases, the ergodic rate degradation becomes more pronounced, since densely obstructed environments increase the probability of severe blockage events. For instance, when the obstacle intensity is set to 0.01 m$^{ - 2}$, the decrease in the ergodic rate is negligible. In contrast, when the obstacle intensity is set to 0.1 m$^{ - 2}$, both PASS and CASS exhibit a much steeper decline in ergodic rate. Additionally, the approximate results for PASS closely track the analytical results and Monte Carlo simulations across the entire range, which demonstrates that the approximate expressions offer an efficient and accurate method for predicting the performance of PASS in the presence of random LoS blockage.

\begin{figure}[t!]
  \centering
  \includegraphics[width= 3.1 in]{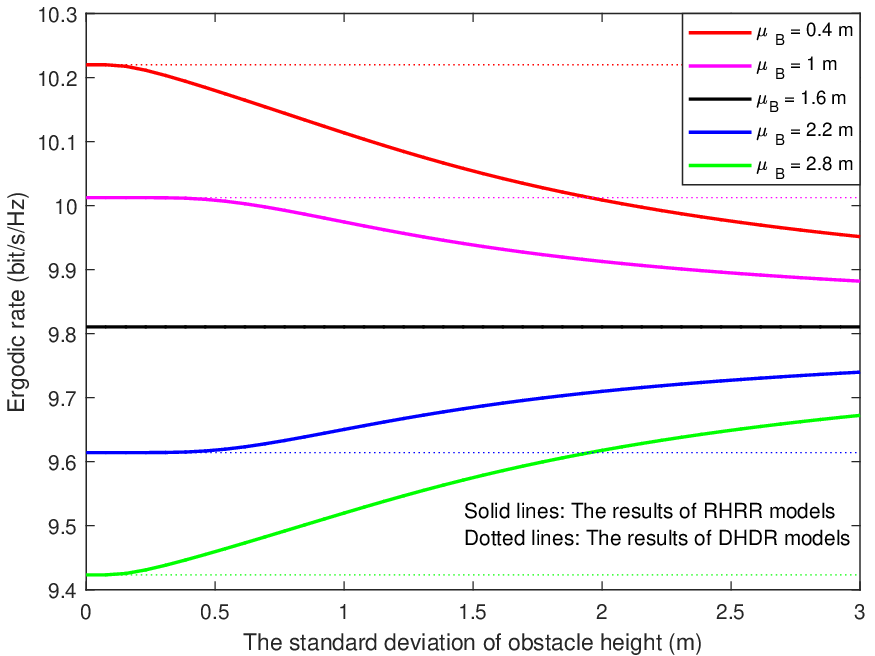}
  \caption{Ergodic rate versus the standard deviation of obstacle heights under the RHRR and DHDR models, respectively.}
  \label{fig:ER3_with_HBstd_compHBmean}
\end{figure}

\emph{8) Ergodic rate affected by different standard deviations of obstacle heights:} Fig.~\ref{fig:ER3_with_HBstd_compHBmean} illustrates the ergodic rate as a function of the standard deviation of obstacle heights under the RHRR and DHDR models. The deterministic obstacle heights in the DHDR models are set to the mean values of the random obstacle heights in the RHRR models. Similarly, the deterministic obstacle radii in the DHDR models are set to the mean values of the random obstacle radii in the RHRR models. Overall, three main conclusions can be drawn. First, the standard deviation of obstacle height reflects the variability of obstructed environments. An increase in the variability of obstacle heights amplifies the randomness of blockage, leading to a gradual convergence of the ergodic rate curves associated with different mean heights. Second, when the mean obstacle height is close to the median height of the LoS path (i.e., 1.6 m), the ergodic rate becomes insensitive to the variance in obstacle height. The deviations above and below the LoS mid-height have symmetrical physical effects on blockage, meaning that the LoS mid-height represents a balance point in blockage events. Therefore, it can be concluded that the performance of PASS can be enhanced by ensuring that the mean obstacle height remains below the LoS mid-height, or equivalently, by elevating the PA to raise the LoS mid-height, thereby reducing the blockage effect of surrounding obstacles. Third, when the standard deviation is sufficiently small (e.g., below 0.5 m), since the obstacle heights cluster tightly around their mean, the results of RHRR models and corresponding DHDR models are nearly identical. It is suggested that the simpler DHDR model could replace the more complex RHRR model without sacrificing accuracy, which reduces analytical complexity and provides a practical tool for performance evaluation.

\section{Conclusions}\label{sec:Concl}
In this work, we investigated the impact of random LoS blockage on PASS by using a stochastic geometry approach. Obstacles were modeled through a homogeneous PPP, which allowed their number and locations to be random across the service area, as the spatial distribution and geometry of obstacles environments were inherently random and irregular in practice. The height and radius of obstacles were random variables in the proposed RHRR model, whereas they were deterministic values in the simplified DHDR model. Closed-form expressions for the LoS probability were derived. Based on the proposed models, the analytical and asymptotic outage probability expressions for randomly located UEs were obtained, together with the analytical and approximate ergodic rate expressions. The proposed blockage models offered a tractable and physically interpretable framework for characterizing LoS blockage in PASS, providing fundamental insights into the design of blockage-resilient PASS in future wireless networks. Future research could extend the analysis of random LoS blockage to more complex scenarios, such as multi-PA and multi-waveguide cooperative scenarios, as well as indoor environments involving both LoS and non-LoS channels.

\numberwithin{equation}{section}
\section*{Appendix~A\\ Proof of Lemma~\ref{lemma:LoS_pr_norm}}\label{Appendix:As}
\renewcommand{\theequation}{A.\arabic{equation}}
\setcounter{equation}{0}
For the uniform random variable representing the radius of obstacles, $R_B$, the probability that the distance from the centre of the obstacles to the LoS path in the top view is less than or equal to the radius of the obstacles can be expressed as follows:
\begin{equation}\label{pr_t lower than RB}
\begin{split}
\mathbb{P} & \left( {\left| t \right| \leqslant {R_B}} \right) = 1 - {F_{R_B}}\left( {\left| t \right|} \right) \\
& = \begin{cases}
    1, & \left| t \right| \leqslant R_{B,\min }, \\
    \frac{{R_{B,\max }} - \left| t \right|}{R_{B,\max }}, & R_{B,\min } < \left| t \right| \leqslant R_{B,\max }, \\
    0, & \left| t \right| > R_{B,\max },
  \end{cases}
\end{split}
\end{equation}
where $F_{R_B}\left( \left| t \right| \right)$ denotes the cumulative distribution function (CDF) of the radius of the obstacles.

The integral of~\eqref{pr_t lower than RB} in~\eqref{expectation_of_blockage_number} can be further written as follows:
\begin{equation}\label{pr_t lower than RB_int}
\begin{aligned}
& \int_{ - \infty }^\infty {\mathbb{P}\left( {\left| t \right| \leqslant {R_B}} \right)dt} \\
& = 2\left( {\int_0^{{R_{B,\min }}} {1dt}  + \int_{{R_{B,\min }}}^{{R_{B,\max }}} {\frac{{{R_{B,\max }} - t}}{{{R_{B,\max }} - {R_{B,\min }}}}dt} } \right) \\
& = {R_{B,\min }} + {R_{B,\max }}.
\end{aligned}
\end{equation}

For the Gaussian random variable representing the height of the obstacles, $H_B$, the probability that an obstacle exceeds the LoS path height is given by:
\begin{equation}\label{pr normHB_exceeds_HLoS}
\mathbb{P}\left( {{H_B} \geqslant {H_{{\mathrm{LoS}}}}\left( l \right)} \right) = 1 - \Phi \left( {\frac{{{H_{{\mathrm{LoS}}}}\left( l \right) - {\mu _B}}}{{{\sigma _B}}}} \right).
\end{equation}

To simplify the calculation, we set $u = {H_{{\mathrm{LoS}}}}\left( l \right)$. The expected number of blockages in~\eqref{expectation_of_blockage_number} can be further written as follows:
\begin{equation}\label{expectation_of_blockage_number_norm}
\begin{aligned}
& \mathbb{E} \left( \Lambda  \right) = \lambda \int_{ - \infty }^\infty  {\mathbb{P}\left( {\left| t \right| \leqslant {R_B}} \right)dt} \int_0^D {\mathbb{P}\left( {{H_B} \geqslant {H_{{\mathrm{LoS}}}}\left( l \right)} \right)dl} \\
& = \frac{{\lambda D\left( {{R_{B,\min }} + {R_{B,\max }}} \right)}}{{H - {H_U}}}\int_{{H_U}}^H {\left[ {1 - \Phi \left( {\frac{{u - {\mu _B}}}{{{\sigma _B}}}} \right)} \right]du}.
\end{aligned} 
\end{equation}

The results in~\eqref{LoS_pr_norm} can be obtained by substituting the expected number of blockages in~\eqref{expectation_of_blockage_number_norm} into the LoS probability in~\eqref{LoS_pr}. Hence, the proof is complete.

\numberwithin{equation}{section}
\section*{Appendix~B\\ Proof of Lemma~\ref{lemma:LoS_pr_constHB}}\label{Appendix:Bs}
\renewcommand{\theequation}{B.\arabic{equation}}
\setcounter{equation}{0}
Based on~\eqref{H_LoS_function}, when the heights of the obstacles can be regarded as a constant, the equivalent condition that an obstacle exceeds the LoS path, i.e., ${{H_B} \geqslant {H_{{\mathrm{LoS}}}}\left( l \right)}$, is given by
\begin{equation}\label{condition_of_HB_bigger_than_HLoS}
l \geqslant {l_0} = D\frac{{H - {H_B}}}{{H - {H_U}}}.
\end{equation}

The probability that the obstacle exceeds the LoS path becomes a step function with respect to the horizontal distance $l$, which can be expressed as:
\begin{equation}\label{pr constHB_exceeds_HLoS}
\mathbb{P}\left( {{H_B} \geqslant {H_{\mathrm{LoS}}}\left( l \right)} \right) =
\begin{cases}
  1, & \mbox{if } l \geqslant l_0, \\
  0, & \mbox{otherwise}.
\end{cases}
\end{equation}

Note that if the height of obstacles is lower than the height of the UE, i.e., ${H_B} < {H_U}$, LoS blockage does not occur, certainly. Thus, without loss of generality, we consider ${H_U} < {H_B} < H$. The integral of~\eqref{pr constHB_exceeds_HLoS} in~\eqref{expectation_of_blockage_number} can be further written as follows:
\begin{equation}\label{Integral_of_pr_HB_bigger_than_HLoS}
\begin{split}
\int_0^D {\mathbb{P}\left( {{H_B} \geqslant {H_{\mathrm{LoS}}}\left( l \right)} \right)dl} & = \int_{\max \left\{ {D,{l_0}} \right\}}^D {1dl} \\
     & = \max \left\{ {0,D - {l_0}} \right\},
\end{split}
\end{equation}
where $\max \left\{ {D,{l_0}} \right\}$ represents the lower limit of the integral, which prevents the probability integral from being negative.

When the radii of the obstacles asymptotically approach their mean, the results in~\eqref{pr_t lower than RB_int} can be approximated as $2R_B$. The expected number of blockages can be further written as follows:
\begin{equation}\label{expectation_of_blockage_number_constHB}
\begin{split}
\mathbb{E}\left( \Lambda  \right) & = 2{R_B}\lambda \left( {D - {l_0}} \right) \\
     & = 2{R_B}\lambda D\frac{H_B - H_U}{H - H_U}.
\end{split}
\end{equation}

By substituting the expected number of blockages in~\eqref{expectation_of_blockage_number_constHB} into the LoS probability in~\eqref{LoS_pr}, the results in~\eqref{LoS_pr_constHB} can be obtained. The proof is complete.

\numberwithin{equation}{section}
\section*{Appendix~C\\ Proof of Theorem~\ref{theorem:OP_norm}}\label{Appendix:Cs}
\renewcommand{\theequation}{C.\arabic{equation}}
\setcounter{equation}{0}
In~\eqref{outage_substitute_PLoS}, the equivalent expressions for the inequality ${\left| {\boldsymbol{\psi} ^{\mathrm{PA}} - \boldsymbol{\psi} } \right| \geqslant \varepsilon }$ are given by
\begin{equation}\label{outage_int_1a}
y \leqslant - \sqrt {\varepsilon ^2 - H_d^2},\quad \mbox{if } y < 0,
\end{equation}
or
\begin{equation}\label{outage_int_1b}
y \geqslant \sqrt {\varepsilon ^2 - H_d^2},\quad \mbox{if } y \geqslant 0.
\end{equation}

By applying the assumption that the UE is uniformly distributed in the service area $\mathcal{A}$, the integrals in~\eqref{outage_substitute_PLoS} can be further written as follows:
\begin{equation}\label{outage_int_3}
\int_{\boldsymbol{\psi}  \in \mathcal{A}} {\left( {1 - {e^{ - \beta D}}} \right)d\boldsymbol{\psi} }  = \frac{1}{W}\int_{ - \frac{W}{2}}^{\frac{W}{2}} {\left( {1 - {e^{ - \beta \left| y \right|}}} \right)dy},
\end{equation}
\begin{equation}\label{outage_int_4}
\begin{gathered}
  \int_{\boldsymbol{\psi}  \in \mathcal{A},\left| {{\boldsymbol{\psi} ^{\mathrm{PA}}} - \boldsymbol{\psi} } \right| \geqslant \varepsilon } {{e^{ - \beta D}}d\boldsymbol{\psi} }  \hfill \\
  \quad \quad \quad \quad  = \frac{1}{W}\int_{ - \frac{W}{2}}^{{\tau _1}} {{e^{ - \beta \left| y \right|}}dy}  + \frac{1}{W}\int_{\tau _2}^{\frac{W}{2}} {{e^{ - \beta \left| y \right|}}dy}  \hfill, \\ 
\end{gathered}
\end{equation}
where $\tau _1$ and $\tau _2$ limit the range of definite integrals to the service area $\mathcal{A}$.
Then, the results in~\eqref{OP norm} can be obtained through some algebraic calculations, and the proof is complete.

\vspace{2.5em}
\numberwithin{equation}{section}
\section*{Appendix~D \\ Proof of Corollary~\ref{coro:ER norm approxi}}\label{Appendix:Ds}
\renewcommand{\theequation}{D.\arabic{equation}}
\setcounter{equation}{0}
By utilizing the symmetry of even functions, the ergodic rate in~\eqref{ER norm} can be expressed as follows:
\begin{equation}\label{Rate_Ergodic_1_AppA}
\bar{\mathcal{R}} = \frac{2}{W\ln 2}\int_0^{\frac{W}{2}} {{e^{ - \beta y}}{g_1}\left( y \right)dy},
\end{equation}
where ${g_1}\left( y \right) = \ln \left( {1 + \frac{\gamma }{y^2 + H_d^2}} \right)$.

By expanding ${g_1}\left( y \right)$ into a Taylor series~\cite{Table_of_integrals}, a closed-form expression of~\eqref{Rate_Ergodic_1_AppA} can be approximately obtained. Considering that the upper limit of the integral, i.e., $W/2$, may not be in the neighborhood of $y=0$, the Taylor series expansion of ${g_1}\left( y \right)$ at $y=a$ is used, which can be written as follows:
\begin{equation}\label{g_1}
{g_1}\left( y \right) \approx {g_1}\left( 0 \right) + {g_1^\prime} \left( 0 \right)y,
\end{equation}
where the first derivative of ${g_1}\left( y \right)$ at $y = a$ is given by
\begin{equation}\label{g_1'}
{g_1^\prime} \left( a \right) =  - \frac{2\gamma }{\ln 2}\frac{a}{\left( a^2 + H_d^2 \right)\left( a^2 + H_d^2 + \gamma \right)}.
\end{equation}

The ergodic rate can be approximated as follows:
\begin{equation}\label{Rate_Ergodic_AppA}
\bar {\mathcal{R}} \approx \frac{2}{W}\left[ {{g_1}\left( a \right){I_0} + {g_1^\prime} \left( a \right){I_1}} \right],
\end{equation}
where
\begin{equation}\label{Rate_Ergodic_AppA_I0}
{I_0} = \int_0^{\frac{W}{2}} {{e^{ - \beta y}}dy}  = \frac{1}{\beta }\left( {1 - {e^{ - \frac{{\beta W}}{2}}}} \right),
\end{equation}
and
\begin{equation}\label{Rate_Ergodic_AppA_I1}
\begin{split}
{I_1} & = \int_0^{\frac{W}{2}} {{e^{ - \beta y}}\left( {y - a} \right)dy} \\
     & = \frac{1}{{{\beta ^2}}}\left[ {1 - {e^{ - \frac{{\beta W}}{2}}}\left( {1 + \frac{{\beta W}}{2}} \right)} \right] - a{I_0}.
\end{split}
\end{equation}

Then, by substituting the average value of the interval $[0, \tfrac{W}{2}]$, which is equal to $\tfrac{W}{4}$, as the center point $y=a$ of the Taylor series expansion, and performing some algebraic operations, the results in~\eqref{ER norm approxi} can be obtained. Hence, the proof is complete.

\numberwithin{equation}{section}
\section*{Appendix~E\\ Proof of Proposition~\ref{propo:high_SNR slope}}\label{Appendix:Es}
\renewcommand{\theequation}{E.\arabic{equation}}
\setcounter{equation}{0}
The high-SNR slope is defined as follows:
\begin{equation}\label{high-SNR slope def}
\mathcal{S} = \mathop {\lim }\limits_{\frac{P}{{{\sigma ^2}}} \to \infty } \frac{{\bar{\mathcal{R}}}}{{{{\log }_2}\left( {1 + \frac{P}{{{\sigma ^2}}}} \right)}}.
\end{equation}

Recall that $\gamma  = \tfrac{\eta P}{\sigma ^2}$ and $\eta = \left( \tfrac{c}{4\pi {f_c}} \right)^2$ are set for the sake of notational simplicity. Thus, the condition ${\tfrac{P}{\sigma ^2} \to \infty }$ is mathematically equivalent to $\gamma \to \infty$. By using L'Hospital's Rule~\cite{sohrab2003basic}, the high-SNR slope can be expressed as follows:
\begin{equation}\label{high-SNR slope proof_1}
\begin{split}
\mathcal{S} & = \mathop {\lim }\limits_{\gamma  \to \infty } \frac{{\frac{{d\bar{\mathcal{R}}}}{{d\gamma }}}}{{\frac{d}{{d\gamma }}{{\log }_2}\gamma }} \\
     & = \mathop {\lim }\limits_{\gamma  \to \infty } \frac{2}{W}\int_0^{\frac{W}{2}} {{e^{ - \beta y}}\frac{\gamma }{y^2 + H_d^2 + \gamma }dy}.
\end{split}
\end{equation}

We set ${g_2}\left( {y,\gamma } \right) = \int_0^{\frac{W}{2}} {e^{ - \beta y} \frac{y^2 + H_d^2}{y^2 + H_d^2 + \gamma }dy}$. For $y \in \left[ {0, \tfrac{W}{2}} \right]$, the following inequality holds true:
\begin{equation}\label{high-SNR slope proof_2}
{g_2}\left( {y,\gamma } \right) \leqslant \int_0^{\frac{W}{2}} {{e^{ - \beta y}}\frac{{{y^2} + H_d^2}}{\gamma }dy}  = \frac{b}{\gamma },
\end{equation}
where $b$ represents a finite constant. For $\gamma \to \infty$, the following equation holds true:
\begin{equation}\label{high-SNR slope proof_3}
\mathop {\lim }\limits_{\gamma  \to \infty } {g_2}\left( {y,\gamma } \right) = \mathop {\lim }\limits_{\gamma  \to \infty } \frac{b}{\gamma } = 0.
\end{equation}

The high-SNR slope can be further expressed as follows:
\begin{equation}\label{high-SNR slope proof_4}
\begin{split}
\mathcal{S} & = \mathop {\lim }\limits_{\gamma  \to \infty } \frac{2}{W}\left[ {\int_0^{\frac{W}{2}} {{e^{ - \beta y}}dy}  - {g_2}\left( {y,\gamma } \right)} \right] \\
     & = \frac{2}{W}\int_0^{\frac{W}{2}} {{e^{ - \beta y}}dy}.
\end{split}
\end{equation}

The results in~\eqref{high-SNR slope} can be obtained after some simple integral manipulation, and the proof is complete.

\bibliographystyle{IEEEtran}
\bibliography{IEEEabrv,LoS_Blockage_PASS}

@ARTICLE{10858129,
  author={You, Changsheng and Cai, Yunlong and Liu, Yuanwei and Di Renzo, Marco and Duman, Tolga M. and Yener, Aylin and Lee Swindlehurst, A.},
  journal={IEEE J. Sel. Areas Commun.}, 
  title={Next Generation Advanced Transceiver Technologies for 6{G} and Beyond}, 
  year={2025},
  volume={43},
  number={3},
  pages={582-627},
  doi={10.1109/JSAC.2025.3536557},
  month={Mar.}
}

@ARTICLE{9140329,
  author={Di Renzo, Marco and Zappone, Alessio and Debbah, Merouane and Alouini, Mohamed-Slim and Yuen, Chau and de Rosny, Julien and Tretyakov, Sergei},
  journal={IEEE J. Sel. Areas Commun.}, 
  title={Smart Radio Environments Empowered by Reconfigurable Intelligent Surfaces: How It Works, State of Research, and The Road Ahead}, 
  year={2020},
  volume={38},
  number={11},
  pages={2450-2525},
  doi={10.1109/JSAC.2020.3007211},
  month={Nov.}
}

@ARTICLE{9424177,
  author={Liu, Yuanwei and Liu, Xiao and Mu, Xidong and Hou, Tianwei and Xu, Jiaqi and Di Renzo, Marco and Al-Dhahir, Naofal},
  journal={IEEE Commun. Surv. Tutorials},
  title={Reconfigurable Intelligent Surfaces: Principles and Opportunities}, 
  year={2021},
  volume={23},
  number={3},
  pages={1546-1577},
  doi={10.1109/COMST.2021.3077737},
  month={thirdquarter}
}

@ARTICLE{10286328,
  author={Zhu, Lipeng and Ma, Wenyan and Zhang, Rui},
  journal={IEEE Commun. Mag.}, 
  title={Movable Antennas for Wireless Communication: Opportunities and Challenges}, 
  year={2024},
  volume={62},
  number={6},
  pages={114-120},
  doi={10.1109/MCOM.001.2300212},
  month={Jun.}
}

@ARTICLE{9264694,
  author={Wong, Kai-Kit and Shojaeifard, Arman and Tong, Kin-Fai and Zhang, Yangyang},
  journal={IEEE Trans. Wireless Commun.}, 
  title={Fluid Antenna Systems}, 
  year={2021},
  volume={20},
  number={3},
  pages={1950-1962},
  doi={10.1109/TWC.2020.3037595},
  month={Mar.}
}

@ARTICLE{suzuki2022pinching,
  author={A. Fukuda and H. Yamamoto and H. Okazaki and Y. Suzuki and K. Kawai},
  journal={NTT DOCOMO Technical J.},
  title={Pinching antenna: Using a dielectric waveguide as an antenna},
  year={2022},
  volume={23},
  number={3},
  pages={5-12},
  month={Jan.}
}

@online{example2025,
  author = {{NTT DOCOMO INC.}},
  title = {Pinching Antenna},
  year = {2022},
  url = {https://www.docomo.ne.jp/english/info/media_center/event/mwc21/pdf/06_MWC2021_docomo_Pinching_Antenna_en.pdf}
}

@ARTICLE{11169486,
  author={Liu, Yuanwei and Wang, Zhaolin and Mu, Xidong and Ouyang, Chongjun and Xu, Xiaoxia and Ding, Zhiguo},
  journal={IEEE Commun. Mag.}, 
  title={Pinching-Antenna Systems: Architecture Designs, Opportunities, and Outlook}, 
  year={2025},
  month = {Early Access,},
  doi={10.1109/MCOM.001.2500037}
}

@ARTICLE{yang2025pinchingantennasprinciplesapplications,
      title={Pinching Antennas: Principles, Applications and Challenges}, 
      author={Zheng Yang and Ning Wang and Yanshi Sun and Zhiguo Ding and Robert Schober and George K. Karagiannidis and Vincent W. S. Wong and Octavia A. Dobre},
      year={2025},
      eprint={2501.10753},
      journal={arXiv:2501.10753},
      primaryClass={cs.IT}
}

@ARTICLE{ding2025flexible,
  author={Ding, Zhiguo and Schober, Robert and Vincent Poor, H.},
  journal={IEEE Trans. Commun.}, 
  title={Flexible-Antenna Systems: A Pinching-Antenna Perspective}, 
  year={2025},  
  month = {Early Access,},
  doi={10.1109/TCOMM.2025.3555866}
}

@ARTICLE{10909665,
  author={Tegos, Sotiris A. and Diamantoulakis, Panagiotis D. and Ding, Zhiguo and Karagiannidis, George K.},
  journal={IEEE Wireless Commun. Lett.}, 
  title={Minimum Data Rate Maximization for Uplink Pinching-Antenna Systems}, 
  year={2025},
  volume={14},
  number={5},
  pages={1516-1520},
  doi={10.1109/LWC.2025.3547956},
  month={May.}
}

@ARTICLE{10912473,
  author={Wang, Kaidi and Ding, Zhiguo and Schober, Robert},
  journal={IEEE Wireless Commun. Lett.}, 
  title={Antenna Activation for {NOMA} Assisted Pinching-Antenna Systems}, 
  year={2025},
  volume={14},
  number={5},
  pages={1526-1530},
  doi={10.1109/LWC.2025.3548280},
  month={May.}
}

@ARTICLE{wang2025modelingbeamformingoptimizationpinchingantenna,
      title={Modeling and Beamforming Optimization for Pinching-Antenna Systems}, 
      author={Zhaolin Wang and Chongjun Ouyang and Xidong Mu and Yuanwei Liu and Zhiguo Ding},
      year={2025},
      eprint={2502.05917},
      journal={arXiv:2502.05917},
      primaryClass={cs.IT}
}

@ARTICLE{11195810,
  author={Hou, Tianwei and Liu, Yuanwei and Nallanathan, Arumugam},
  journal={IEEE Trans. Commun.}, 
  title={On the Performance of Uplink Pinching Antenna Systems ({PASS})}, 
  year={2025},
  month={Early Access,},
  doi={10.1109/TCOMM.2025.3618726}
}

@ARTICLE{10896748,
  author={Xu, Yanqing and Ding, Zhiguo and Karagiannidis, George K.},
  journal={IEEE Wireless Commun. Lett.}, 
  title={Rate Maximization for Downlink Pinching-Antenna Systems}, 
  year={2025},
  volume={14},
  number={5},
  pages={1431-1435},
  doi={10.1109/LWC.2025.3543889},
  month={May.}
}

@ARTICLE{11197530,
  author={Zhang, Zheng and Wang, Zhaolin and Mu, Xidong and He, Bingtao and Chen, Jian and Liu, Yuanwei},
  journal={IEEE Commun. Lett.}, 
  title={Integrated Sensing and Communications for Pinching-Antenna Systems ({PASS})}, 
  year={2025},
  month={Early Access,},
  doi={10.1109/LCOMM.2025.3619778}
}

@ARTICLE{li2025pinchingantennaintegratedsensing,
      title={Pinching Antenna System for Integrated Sensing and Communications}, 
      author={Haochen Li and Ruikang Zhong and Jiayi Lei and Yuanwei Liu},
      year={2025},
      eprint={2508.19540},
      journal={arXiv:2508.19540},
      primaryClass={eess.SP}      
}

@ARTICLE{zhang2025pinchingantennasystemspassbasedindoor,
      title={Pinching-Antenna Systems ({PASS})-based Indoor Positioning}, 
      author={Yaoyu Zhang and Xin Sun and Jun Wang and Tianwei Hou and Anna Li and Yuanwei Liu and Arumugam Nallanathan},
      year={2025},
      eprint={2508.08185},
      journal={arXiv:2508.08185},
      primaryClass={eess.SY}
}

@ARTICLE{11106459,
  author={Li, Yixuan and Wang, Ji and Liu, Yuanwei and Ding, Zhiguo},
  journal={IEEE Commun. Lett.}, 
  title={Pinching-Antenna Assisted Simultaneous Wireless Information and Power Transfer}, 
  year={2025},
  volume={29},
  number={10},
  pages={2341-2345},
  doi={10.1109/LCOMM.2025.3594663},
  month={Oct.}
}

@ARTICLE{6840343,
  author={Bai, Tianyang and Vaze, Rahul and Heath, Robert W.},
  journal={IEEE Trans. Wireless Commun.}, 
  title={Analysis of Blockage Effects on Urban Cellular Networks}, 
  year={2014},
  volume={13},
  number={9},
  pages={5070-5083},
  doi={10.1109/TWC.2014.2331971},
  month={Sep.}
}

@ARTICLE{10121509,
  author={Miao, Haiyang and Zhang, Jianhua and Tang, Pan and Tian, Lei and Zhao, Xinyu and Guo, Bolun and Liu, Guangyi},
  journal={IEEE J. Sel. Areas Commun.}, 
  title={Sub-6 {GHz} to {mmWave} for 5{G}-Advanced and Beyond: Channel Measurements, Characteristics and Impact on System Performance}, 
  year={2023},
  volume={41},
  number={6},
  pages={1945-1960},
  doi={10.1109/JSAC.2023.3274175},
  month={Jun.}
}

@ARTICLE{8869705,
  author={Saad, Walid and Bennis, Mehdi and Chen, Mingzhe},
  journal={IEEE Network}, 
  title={A Vision of 6{G} Wireless Systems: Applications, Trends, Technologies, and Open Research Problems}, 
  year={2020},
  volume={34},
  number={3},
  pages={134-142},
  doi={10.1109/MNET.001.1900287},
  month={May.}
}

@ARTICLE{9512383,
  author={Charan, Gouranga and Alrabeiah, Muhammad and Alkhateeb, Ahmed},
  journal={IEEE Trans. Veh. Technol.}, 
  title={Vision-Aided 6{G} Wireless Communications: Blockage Prediction and Proactive Handoff}, 
  year={2021},
  volume={70},
  number={10},
  pages={10193-10208},
  doi={10.1109/TVT.2021.3104219},
  month={Oct.}
}

@ARTICLE{7593259,
  author={Andrews, Jeffrey G. and Bai, Tianyang and Kulkarni, Mandar N. and Alkhateeb, Ahmed and Gupta, Abhishek K. and Heath, Robert W.},
  journal={IEEE Trans. Commun.}, 
  title={Modeling and Analyzing Millimeter Wave Cellular Systems}, 
  year={2017},
  volume={65},
  number={1},
  pages={403-430},
  doi={10.1109/TCOMM.2016.2618794},
  month={Jan.}
}

@ARTICLE{11036558,
  author={Ding, Zhiguo and Vincent Poor, H.},
  journal={IEEE Wireless Commun. Lett.}, 
  title={{LoS} Blockage in Pinching-Antenna Systems: Curse or Blessing?}, 
  year={2025},
  volume={14},
  number={9},
  pages={2798-2802},
  doi={10.1109/LWC.2025.3579616},
  month={Sep.}
}

@ARTICLE{11178241,
  author       = {Wang, Kaidi and Ouyang, Chongjun and Liu, Yuanwei and Ding, Zhiguo},
  journal      = {IEEE Wireless Commun. Lett.},
  title        = {Pinching-Antenna Systems With {LoS} Blockages},
  year         = {2025},
  month        = {Early Access,},
  doi          = {10.1109/LWC.2025.3614451}
}

@ARTICLE{xu2025pinchingantennadesignlosblockage,
      title={Pinching-Antenna System Design with {LoS} Blockage: Does In-Waveguide Attenuation Matter?}, 
      author={Yanqing Xu and Zhiguo Ding and Octavia A. Dobre and Tsung-Hui Chang},
      year={2025},
      eprint={2508.07131},
      journal={arXiv:2508.07131},
      primaryClass={eess.SP}
}

@ARTICLE{9516701,
  author={Lu, Xiao and Salehi, Mohammad and Haenggi, Martin and Hossain, Ekram and Jiang, Hai},
  journal={IEEE Commun. Surv. Tutorials}, 
  title={Stochastic Geometry Analysis of Spatial-Temporal Performance in Wireless Networks: A Tutorial}, 
  year={2021},
  volume={23},
  number={4},
  pages={2753-2801},
  doi={10.1109/COMST.2021.3104581},
  month={Fourthquarter}
}

@INPROCEEDINGS{human_blockage1,
  author={Gapeyenko, Margarita and Samuylov, Andrey and Gerasimenko, Mikhail and Moltchanov, Dmitri and Singh, Sarabjot and Aryafar, Ehsan and Yeh, Shu-ping and Himayat, Nageen and Andreev, Sergey and Koucheryavy, Yevgeni},
  booktitle={Proc. IEEE Int. Conf. Commun. (ICC)}, 
  title={Analysis of human-body blockage in urban millimeter-wave cellular communications}, 
  year={2016},
  volume={},
  number={},
  pages={1-7},
  address = {Kuala Lumpur, Malaysia},
  doi={10.1109/ICC.2016.7511572},
  month={May.}
}

@ARTICLE{9253591,
  author={Ruiz, Cristian García and Pascual-Iserte, Antonio and Muñoz, Olga},
  journal={IEEE Trans. Veh. Technol.}, 
  title={Analysis of Blocking in {mmWave} Cellular Systems: Characterization of the {LOS} and {NLOS} Intervals in Urban Scenarios}, 
  year={2020},
  volume={69},
  number={12},
  pages={16247-16252},
  doi={10.1109/TVT.2020.3037125},
  month={Dec.}
}

@ARTICLE{8643739,
  author={Jain, Ish Kumar and Kumar, Rajeev and Panwar, Shivendra S.},
  journal={IEEE J. Sel. Areas Commun.}, 
  title={The Impact of Mobile Blockers on Millimeter Wave Cellular Systems}, 
  year={2019},
  volume={37},
  number={4},
  pages={854-868},
  doi={10.1109/JSAC.2019.2898756},
  month={Apr.}
}

@ARTICLE{9492764,
  author={Chen, Chen and Zhang, Jiliang and Chu, Xiaoli and Zhang, Jie},
  journal={IEEE Trans. Veh. Technol.}, 
  title={On the Optimal Base-Station Height in mmWave Small-Cell Networks Considering Cylindrical Blockage Effects}, 
  year={2021},
  volume={70},
  number={9},
  pages={9588-9592},
  doi={10.1109/TVT.2021.3098626},
  month = {Sep.}
}

@ARTICLE{9247469,
  author={Wu, Yongzhi and Kokkoniemi, Joonas and Han, Chong and Juntti, Markku},
  journal={IEEE Trans. Wireless Commun.}, 
  title={Interference and Coverage Analysis for Terahertz Networks With Indoor Blockage Effects and {Line-of-Sight} Access Point Association}, 
  year={2021},
  volume={20},
  number={3},
  pages={1472-1486},
  doi={10.1109/TWC.2020.3033825},
  month={Mar.}
}

@techreport{3gpp.38.901,
 author = {3rd Generation Partnership Project (3GPP)},  
 title = {Study on channel model for frequencies from 0.5 to 100 {GHz} ({Release} 19)}, 
 number = {38.901}, 
 type = {Tech. Rep.}, 
 year = {2025},
 month = {Jul.}
}

@book{chiu2013stochastic,
  title={Stochastic geometry and its applications},
  author={Chiu, Sung Nok and Stoyan, Dietrich and Kendall, Wilfrid S and Mecke, Joseph},
  year={2013},
  edition = {3rd},
  publisher={John Wiley \& Sons Inc},
  address={Chichester, West Sussex, U.K.}
}

@book{sohrab2003basic,
  title={Basic real analysis},
  publisher={Springer},  
  address={New York, NY, USA},
  year={2003},
  author={Sohrab, Houshang H}
}

@BOOK{Table_of_integrals,
  title = {Table of Integrals, Series and Products},
  publisher = {New York: Academic Press},
  year = {2000},
  author = {I. S. Gradshteyn and I. M. Ryzhik},
  edition = {6th},
}

\end{document}